\documentclass[aps,pre,showpacs,floats,twocolumn,floats,superscriptaddress,floatfix]{revtex4-1}
\usepackage{bm}
\usepackage{verbatim}
\usepackage{graphicx}
\usepackage{graphics,epsfig}
\usepackage{theorem}
\usepackage{makeidx}
\usepackage{amsmath}
\usepackage{epic}
\usepackage{amscd}
\usepackage{xy}
\usepackage{amssymb}
\usepackage{epsfig}
\begin{document}
\newif\iffigs 
\figstrue
\iffigs \fi
\def\drawing #1 #2 #3 {
\begin{center}
\setlength{\unitlength}{1mm}
\begin{picture}(#1,#2)(0,0)
\put(0,0){\framebox(#1,#2){#3}}
\end{picture}
\end{center} }

\newcommand{\ul}{{\bm u}_{\scriptscriptstyle \mathrm{L}}}
\newcommand{\deltal}{{\delta}_{\scriptscriptstyle \mathrm{L}}}
\newcommand{\ue}{\mathrm{e}}
\newcommand{\ui}{\mathrm{i}\,}
\newcommand{\kg}{{k_{\scriptscriptstyle \mathrm{G}}}}
\def\v{\bm v}
\def\x{\bm x}
\def\k{\bm k}
\def\ds{\displaystyle}

\title{Statistics of the inverse-cascade regime in
two-dimensional magnetohydrodynamic turbulence}
\author{Debarghya Banerjee}
\email{debarghya@physics.iisc.ernet.in}
\affiliation{Centre for Condensed Matter Theory,
Department of Physics, Indian Institute of Science,
Bangalore 560012, India}
\author{Rahul Pandit}
\email{rahul@physics.iisc.ernet.in}
\altaffiliation[Also at ]{Jawaharlal Nehru Centre For Advanced
Scientific Research, Jakkur, Bangalore, India}
\affiliation{Centre for Condensed Matter Theory,
Department of Physics, Indian Institute of Science,
Bangalore 560012, India}

\begin{abstract}
We present a detailed direct numerical simulation of
statistically steady, homogeneous, isotropic, two-dimensional
magnetohydrodynamic (2D MHD) turbulence. Our study concentrates
on the inverse cascade of the magnetic vector potential. We
examine the dependence of the statistical properties of such
turbulence on dissipation and friction coefficients. We
extend earlier work significantly by calculating fluid and
magnetic spectra, probability distribution functions (PDFs) of
the velocity, magnetic, vorticity, current, stream-function, and
magnetic-vector-potential fields and their increments. We
quantify the deviations of these PDFs from Gaussian ones by
computing their flatnesses and hyperflatnesses. We also present
PDFs of the Okubo-Weiss parameter, which distinguishes between
vortical and extensional flow regions, and its magnetic analog.
We show that the hyperflatnesses of PDFs of the increments of the
stream-function and the magnetic vector potential exhibit
significant scale dependence and we examine the implication of
this for the multiscaling of structure functions.  We compare
our results with those of earlier studies.
\end{abstract}

\keywords{MHD, turbulence, multiscaling , inverse cascade}

\date{\today}
\pacs{47.27.-i, 47.27.ek, 47.27.er, 47.27.Gs}

\maketitle

\section{Introduction}

The energy spectrum $E(k)$ describes the distribution of energy over the
wave-number ($k$) scales in a turbulent fluid; it is, therefore, an important
statistical measure of the characteristics of fluid turbulence. In
three-dimensional (3D) fluid turbulence, energy, injected at $k=k_{\rm inj}\equiv 2
\pi/l_{\rm inj}$, cascades down to small length scales ($k > k_{\rm inj}$), because of
the nonlinearities in the system, and leads to the scaling form $E(k) \sim
k^{-\alpha}$, in the inertial range $k_{\rm inj} \ll k \ll k_d$, where $k_d = 2
\pi/\eta_d$ and the Kolmogorov dissipation length scale is $\eta_d$; at the level
of Kolmogorov's~\cite{K41,frisch95} phenomenological theory (K41), $\alpha =
-5/3$. This Richardson \textit{forward} cascade of energy, from large to small
length scales, continues until the length scale $\eta_d$, at which point
viscous dissipation becomes significant~\cite{frisch95,richardson} and $E(k)$
falls very rapidly for $k > k_d$.

In two-dimensional (2D), statistically steady turbulence, the
Richardson forward cascade is replaced by two cascades, one
forward and the other
inverse~\cite{frisch95,fjortoft,kraichnan,leith,batchelor,lesieurbook,boffettaecke12,pandit09},
because of the conservation of energy and enstrophy in the
unforced, inviscid limit. In the forward-cascade regime,
enstrophy cascades from $k_{\rm inj}$ to larger values of $k$ and
leads to $E(k) \sim k^{-\delta}$, where $\delta$ depends on the
friction on the 2D fluid film (if there is no friction, $\delta =
3$); by contrast, in the inverse-cascade regime, energy cascades
from $k_{\rm inj}$ to smaller values of $k$, which leads to the
formation of large vortices, whose size is limited, finally, by
the
friction~\cite{lesieurbook,boffettaecke12,pandit09,rutgers,perlekarnjp,mitra11}.
Direct numerical simulations (DNSs) have been used to study the
two cascades in 2D fluid turbulence in great
detail~\cite{boffetta}.

Inverse energy cascades have been found and investigated in other
turbulent systems, such as quasi-geostrophic
flows~\cite{boffettanjp,falkovich} and turbulence in fluid films with
polymer additives~\cite{anupampaper}. It has also been noted that
quantities other than the energy can show inverse cascades;
examples include the inverse cascade of magnetic helicity in three-dimensional (3D)
MHD turbulence~\cite{muller,alexakis} and its analog
in 2D MHD turbulence~\cite{biskamp}. The 3D MHD equations
conserve the energy $E$, cross helicity, and magnetic helicity in
the inviscid, unforced case; when viscosity and magnetic
diffusivity are included, forced 3D MHD turbulence displays a
nonequilibrium statistically steady state in which the magnetic
helicity displays an inverse cascade, from the forcing length
scale to larger scales, whereas the energy and cross helicity
display forward cascades~\cite{pouquet,muller}. The 2D MHD analog
of the  magnetic helicity is $|\psi|^2$, where $\psi$ is the 2D
scalar analog of the 3D magnetic vector
potential~\cite{pouquet78}. The inverse cascade in turbulent MHD
is very important in the formation of large-scale structures in
astrophysical plasmas~\cite{alexakis,mhd}.

To provide some background for our study, we begin with 
dimensional predictions~\cite{biskamp,celani} for the fluid- and 
magnetic-energy spectra, $E^u(k)$ and $E^b(k)$, respectively, in the
2D MHD inverse-cascade regime, which we have described above. 
The dimensions (indicated by square brackets and expressed as
powers of length $L$ and time $T$) of various quantities are given 
below (velocity and magnetic fields have the same
units in the standard formulation of MHD~\cite{biskampbook,mkvreview}):
\begin{eqnarray}
\lbrack \psi \rbrack &=& \frac{L^2}{T} ;\, 
\lbrack \phi \rbrack = \frac{L^2}{T} ; \nonumber \\
\lbrack k \rbrack &=& \frac{1}{L}  ;\nonumber \\
\lbrack \epsilon_{\psi} \rbrack &=& \frac{L^4}{T^3} ;\,
\lbrack \epsilon_{\phi} \rbrack = \frac{L^4}{T^3} ; \nonumber \\
\lbrack |\psi(k)|^2 \rbrack &=& \frac{L^5}{T^2} ;\,
\lbrack |\phi(k)|^2 \rbrack = \frac{L^5}{T^2}; 
\label{eq:dimensions1}
\end{eqnarray}
here, $\epsilon_{\psi}$ and $\epsilon_{\phi}$ are, respectively,
dissipation rates per unit time for $|\psi|^2$ and $|\phi|^2$, 
and the arguments $k$ denote spatial fourier transforms.
$|\psi|^2$ displays an inverse cascade, so we can make a 
Kolmogorov-type ansatz~\cite{biskampbook,K41,frisch95,lesieurbook} 
for its spectrum, namely, in the inverse-cascade region,
$\epsilon_{\psi}$ should not depend on viscosity, magnetic
diffusivity, and friction. Therefore, we write
\begin{equation}
|\psi(k)|^2 \sim
\epsilon_{\psi}^{\gamma_1} k^{\gamma_2};
\label{eq:dimensions2}
\end{equation}
and then dimensional analysis (Eq.~\ref{eq:dimensions1}) requires
\begin{equation}
\frac{L^5}{T^2} = \left( \frac{L^4}{T^3} \right)^{\gamma_1} 
\left( \frac{1}{L} \right)^{\gamma_2}, 
\label{eq:dimensions3}
\end{equation}
from which we find $\gamma_1 = 2/3$ and  $\gamma_2 = -7/3$ and
$|\psi(k)|^2 \sim k^{-7/3}$, i.e., 
\begin{equation}
E^b(k)\sim k^2 |\psi(k)|^2 \sim k^{-1/3}.
\label{eq:dimensions4}
\end{equation} 

To obtain the scaling form for  $E^u(k)$, we must
make an additional assumption. One such assumption~\cite{celani}
is that the nonlinear terms in the velocity equation in 2D MHD
(see below) must balance each other in the inverse-cascade
regime; this assumption implies $E^u(k)\sim E^b(k)\sim k^{-1/3}$.
It has been suggested~\cite{biskamp} that this result is valid
only in the forward-cascade parts of these energy spectra,
because the large-scale part of the magnetic field leads to an
Alfv\'en-type effect, which yields, in turn, a strong coupling between
small-scale velocity and magnetic fields and implies, thereby,
that  $E^u(k)\sim E^b(k)$; Ref.~\cite{biskamp} goes on to suggest
that this effect may not operate in the inverse-cascade regime in 
2D MHD. Our study has been designed to explore, among other
issues, the scaling forms of such spectra in this regime.

We build on earlier studies of inverse cascades in
homogeneous, isotropic, and statistically steady MHD turbulence,
in both 3D~\cite{muller} and 2D~\cite{biskamp}, by carrying out
extensive direct numerical simulations (DNSs) of the forced 2D
MHD equations, with friction, and forcing that yields
a substantial spectral regime in which there is a substantial
inverse-cascade region in the spectrum for $|\psi|^2$. We investigate two
cases R1, with a finite, positive friction,  and R2, with zero
friction.  Our study yields a variety of interesting and
unforeseen results. 

The most remarkable result of our study is that the runs R1 and
R2 yield quantitatively different statistical properties. In
particular, run R1 yields fluid- and magnetic-energy spectra
whose scaling forms in the inverse-cascade regime are,
respectively, consistent with $E^u(k) \sim k^0$ and $E^b(k) \sim
k^{-1/3}$; by contrast, run R2 yields energy spectra that are
consistent with $E^u(k) \sim k^{1/3}$ and $E^b(k) \sim k^{-1/3}$.
Furthermore, we find that, in both runs R1 and R2, the PDFs of
$\omega$, $j$, $\psi$, and $\phi$ (the fluid stream function)
are close to Gaussian; we characterize mild deviations from
Gaussian forms by calculating the kurtoses for these PDFs.  The
PDFs of field increments, such as $\delta \omega = \omega({\bf r}
+ {\bf l}) - \omega({\bf r})$, are also predominantly Gaussian,
but their hyperflatnesses $F_6(l)$, which depend on the length
scale $l = |{\bf l}|$, show deviations from the Gaussian value of
15, especially for the increments $\psi$ and $\phi$. The angle
between $\omega$ and $j$ is $\beta_{\omega,j}$ and that between
${\bf u}$ and ${\bf b}$ is $\beta_{u,b}$. We obtain the PDFs of
these angles and show therefrom that, in both runs R1 and R2, (a)
$\omega$ and $j$ are perfectly aligned or anti-aligned, with
equal probability, and (b) the PDF of $\cos(\beta_{u,b})$ is
symmetrical about a minima at $\beta_{u,b} = 0$ and attains its highest
values at $\beta_{u,b} = \pm \pi$. We obtain
the Okubo-Weiss parameter $\Lambda$, which is positive (negative)
in regions where the fluid flow is vortical (extensional), and
its magnetic analog $\Lambda_b$; we then obtain PDFs of 
$\Lambda$ and $\Lambda_b$ and also their joint PDF; and we
show that they are qualitatively similar for runs R1 and R2;
in particular, the PDFs have a cusp at the origin and tails that 
we fit to exponential forms. We also explore the scaling and
possible multiscaling of structure functions of field increments.

The remaining part of this paper is organized as follows. Section
II contains a description of equations of 2D MHD, the quantities
we calculate, and the numerical methods we use. In Section III we
present our results. Section IV contains our conclusions and a
discussion of our results.

\section{model and numerical methods}

The 2D MHD equations can be written as
\begin{eqnarray}
\nonumber
\frac{\partial \omega}{\partial t} + {\bf u} \cdot {\bf \nabla} \omega + \mu^{\omega} \omega &=& -\nu(-\nabla^2)^{\alpha} \omega + f^{\omega} + {\bf b} \cdot {\bf \nabla} j, \\
\nonumber
\frac{\partial \psi}{\partial t} + {\bf u} \cdot {\bf \nabla} \psi + \mu^{\psi} \psi &=& -\eta(-\nabla^2)^{\alpha} \psi + f^{\psi}, \\
\label{eq:2dmhd}
\end{eqnarray} 
where $\bf b$, the magnetic field, and $\bf u$, the velocity
field, are related to the 2D analog of the magnetic vector
potential $\psi$ and the stream function $\phi$ as follows: ${\bf
b} = \hat{z} \times {\bf \nabla} \psi$ and ${\bf u} = \hat{z}
\times {\bf \nabla} \phi$, where $\hat{z}$ is the unit vector
normal to our 2D simulation domain. The current density and
vorticity fields are, respectively, $j = \nabla^2 \psi$ and
$\omega = \nabla^2 \phi$. In this form, the 2D MHD equations
satisfy the incompressibility condition $\nabla \cdot {\bf u}$ 
$=0$ and $\nabla \cdot \bf{ b}$ $= 0$. The order of the
dissipativity is $\alpha$, the coefficients of friction in the 
two equations are  $\mu^{\omega}$ and $\mu^{\psi}$, respectively,
and $\nu$ and $\eta$ are, respectively, the kinematic viscosity
and magnetic diffusivity; $f^{\omega}$ and $f^{\psi}$ are the forcing terms. 
We carry out two DNSs of these equations; the first DNS (R1) has
hyperviscosity and hyperdiffusivity with $\alpha =2$, which allow
us to obtain large inertial ranges in energy spectra; and the
second DNS (R2) has conventional viscosity and diffusivity with
$\alpha =1$; the other parameters for these runs are given in
Table~\ref{table:param}. In the DNS R2,
$\mu^{\omega}=\mu^{\psi}=0$, so the inverse cascade in $\psi$
leads to an accumulation of magnetic energy at small $k$ in the
magnetic-energy spectrum (Fig.~\ref{fig:spectra}).  To control this small-$k$
accumulation and to obtain a statistically steady state, we use
$\mu^{\omega}=0$ but $\mu^{\psi} > 0$ in DNS R1. Our DNSs use a
standard pseudo-spectral method~\cite{canuto,biskamp,perlekarnjp} in a
two-dimensional, square simulation domain with side $\mathbb{L}=2 \pi$ and
periodic boundary conditions in both $x$ and $y$ directions; we
remove aliasing errors by using a two-third dealiasing method.
For time marching, we use a second-order, Runge-Kutta method.

\begin{table*}[htbp]
\begin{center}
\begin{tabular}{|l|c|c|c|c|c|c|c|c|c|c|c|c|c|c|c|}
\hline
Runs & $N$ & $\nu=\eta$ & $\alpha$ & $\mu^{\omega}$ & $\mu^{\psi}$ & $k_{\rm inj}$ & $f_{\rm amp}^{\omega}$ & $f_{\rm amp}^{\psi}$ & $k_{\rm max}/k_{\rm inj}$ & $\nu_{\rm eff}$ & $\eta_{\rm eff}$ & $Re$ & $Re_M$ & $\tau_{\rm eddy}$ & $\tau_{\rm av}/\tau_{\rm eddy}$ \\ \hline

R1 & $4096$ & $10^{-9}$ & $2$ & $0$ & $5.0 \times 10^{-5}$ & $500$ & $10^{-2} $ & $10^{-3}$ & $2.7$ & $2.54 \times 10^{-4}$ & $3.06 \times 10^{-4}$ & $21796$ & $18092$ & $7$ & $30$\\

R2 & $1024$ & $10^{-3}$ & $1$ & $0$ & $0$ & $70$ & $0$ & $10^{-3}$ & $4.9$ & $10^{-3}$ & $10^{-3}$ & $1362$ & $1362$ & $30$ & $33$\\ \hline
\end{tabular}
\end{center}
\caption{The values of the different parameters used in our runs R1 and R2 with 
$N^2$ collocation points. $\nu$ and $\eta$ are, respectively, the kinematic viscosity
and magnetic diffusivity, $\alpha$ is the order of the dissipativity, the coefficients 
of friction in the two equations are  $\mu^{\omega}$ and $\mu^{\psi}$, 
$k_{\rm inj}$ is the energy injection scale, the forcing terms are $f^{\omega}$ and
$f^{\psi}$, $k_{\rm max}$ is the largest resolved wavenumber, and the ratio 
$k_{\rm max}/k_{rm inj}$ gives us an idea whether the simulation resolves small scales
sufficiently, $\nu_{\rm eff}$ and $\eta_{\rm eff}$ gives the effective kinematic 
viscosity and magnetic diffusivity respectively, $Re$ and $Re_{M}$ gives the fluid
and magnetic Reynolds number respectively, $\tau_{\rm eddy}$ is the eddy turnover time,
and $\tau_{\rm av}$ is the time over which we average our data.}
\label{table:param}
\end{table*}

We obtain a statistically steady state by the following
procedure: We first carry out a DNS of Eqs.~(\ref{eq:2dmhd}) with
$f^{\omega}=f^{\psi}=0$, i.e., no forcing, and an initial condition
$\phi(k) = e^{-2 k^2 + \iota \theta_1}$, $\psi(k) = e^{-2 k^2 + \iota \theta_2}$, 
where $\theta_1$ and $\theta_2$ are independent random phases 
distributed uniformly on the interval $[0, 2\pi]$.
We then evolve the system until we obtain a peak 
in the energy-dissipation rate; this signals that the 
energy in the low-$k$ modes has cascaded down to high-$k$ modes.
At this time, we start forcing the system at a large value of 
$k=k_{\rm inj}$, in order to obtain a clear, inverse-cascade
regime. We use the following forcing terms:
\begin{eqnarray}
f^{\omega} &=& -f_{\rm amp}^{\omega} k_{\rm inj} \cos(k_{\rm inj} x);
\nonumber \\
f^{\psi} &=& f_{\rm amp}^{\psi} \frac{1}{k_{\rm inj}} \cos(k_{\rm inj} y).
\label{eq:force}
\end{eqnarray}
We now allow the system to reach a statistically steady state,
maintain it in this state for $\simeq 10 \tau_{\rm eddy}$,
where $\tau_{\rm eddy}$ is the box-size eddy-turnover time, and then
collect data for an averaging time $\tau_{\rm av}$ (see
Table~\ref{table:param}) for the statistical properties we study.

In addition to the spatiotemporal evolution of $\omega$ and
$\psi$, we obtain $\bf{u}, \, \bf{b}, \, \phi$, and $j$.  The fluid
Reynolds number is  $Re = v_{\rm rms} 2 \pi/\nu_{\rm eff}$, its 
magnetic analog is $Re_M =
v_{\rm rms} 2 \pi/\eta_{\rm eff}$, the root-mean-square velocity is 
$v_{\rm rms} = \sqrt{E^u}$, the effective
viscosity and magnetic diffusivity are, respectively,
\begin{eqnarray}
\nu_{\rm eff} &=& \frac{\sum_k \nu k^{2\alpha} E^u(k)}{\sum_k  k^{2} E^u(k)} , \nonumber \\
\eta_{\rm eff} &=& \frac{\sum_k \eta k^{2\alpha} E^b(k)}{\sum_k  k^{2} E^b(k)},
\end{eqnarray}
the box-size eddy turnover time is $\tau_{\rm eddy} = 2\pi/v_{\rm rms}$,
and the kinetic- and magnetic-energy spectra are
$E^u(k) = \Sigma_{{\bf k} \ni |{\bf k}| =k}
|{\bf u (k)}|^2$ and $E^b(k) = \Sigma_{{\bf k} \ni |{\bf k}| =k} |{\bf b
(k)}|^2$, respectively. We also calculate the
fluid Okubo-Weiss parameter~\cite{okubo,weiss,perlekarnjp} 
\begin{equation}
\Lambda = -\left( \frac{\partial u_x}{\partial x} \right )^2 - \frac{\partial
u_y}{\partial x} \frac{\partial u_x}{\partial y}. 
\label{eq:lamb1}
\end{equation}
For a fluid, in the inviscid, unforced case without friction, the sign of
$\Lambda$ can be used to distinguish between vortical and extensional regions
of the flow. In particular, the flow is vortical, if $\Lambda > 0$, and it is
extensional, if $\Lambda <0$. This criterion works well even in the presence
of viscosity, friction, and forcing~\cite{perlekarnjp}. 

It is useful to introduce the magnetic analog $\Lambda_b$ of 
the fluid Okubo-Weiss parameter $\Lambda$;  $\Lambda_b$ follows from the 
determinant of the magnetic-field-gradient 
tensor and can be written as the difference of the
squares of the current density and the magnetic strain
rate~\cite{shivammogi}. Thus, we expect that $\Lambda_b > 0 $ in 
current-dominated regions, whereas $\Lambda_b < 0$ in regions
that are dominated by the magnetic strain rate. Specifically,
$\Lambda_b$ is defined as follows~\cite{shivammogi}:
\begin{equation}
\Lambda_b = -\left( \frac{\partial b_x}{\partial x} \right )^2 - \frac{\partial 
b_y}{\partial x} \frac{\partial b_x}{\partial y}.
\label{eq:lamb2}
\end{equation}

We obtain field-increment PDFs and structure functions from
$\delta \omega = \omega({\bf r} + {\bf l}) - \omega({\bf r})$ and
similar equations for $\delta j$, $\delta \psi$, and $\delta \phi
$.  We also calculate similar PDFs for the longitudinal
components of the velocity- and magnetic-field increments, i.e.,
$\delta u_{||} = ({\bf u}({\bf r} + {\bf l}) - {\bf u}({\bf
r}))\cdot {\bf l }/l$ and its magnetic counterpart. The
order-$p$, longitudinal structure function of a field $a$ is
$S_p^a(l)=\langle (\delta a(l))^p \rangle$; when we calculate
$S_p^a(l)$, we subtract the mean-flow value of $a$ from its value
at a given point in space, average  $S_p^a(l)$ over (i) a circle
of radius $l$, (ii) different, representative origins ${\bf r}$,
and (iii) over $\simeq 30$ independent field configurations,
which are separated from each other by one $\tau_{\rm eddy}$.

\section{results}

We present our results in three subsections. The first subsection
(\ref{section_spectra}) is devoted to a discussion of the spectra
we obtain from runs R1 and R2. The second subsection
(\ref{section_onepoint}) deals with one-point statistics and
PDFs. The third subsection (\ref{section_twopoint}) contains
two-point statistics, such as PDFs of field increments and
structure functions.

\subsection{Spectra}
\label{section_spectra}

\begin{figure*}[htbp]
\includegraphics[width=8cm,height=5cm]{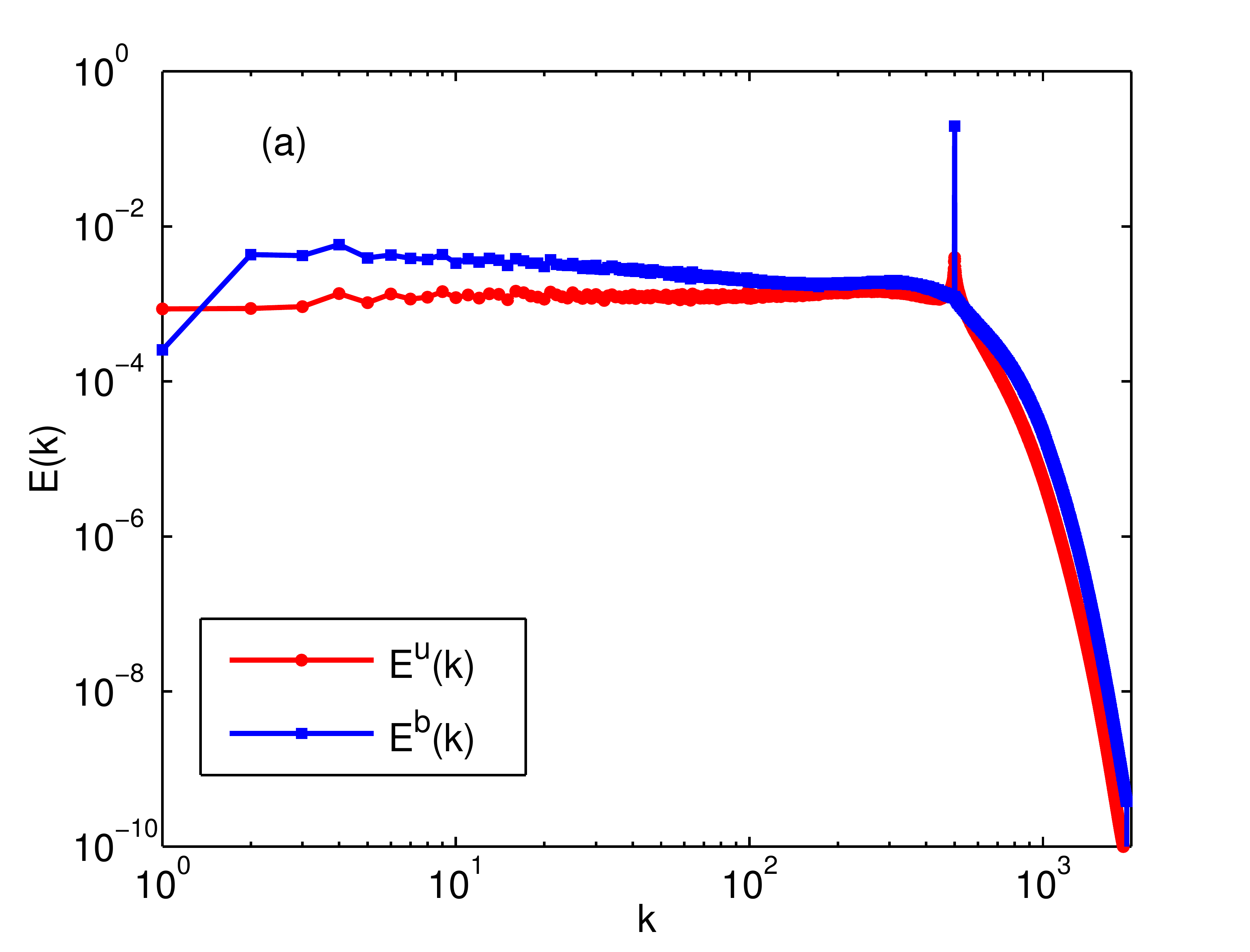}
\includegraphics[width=8cm,height=5cm]{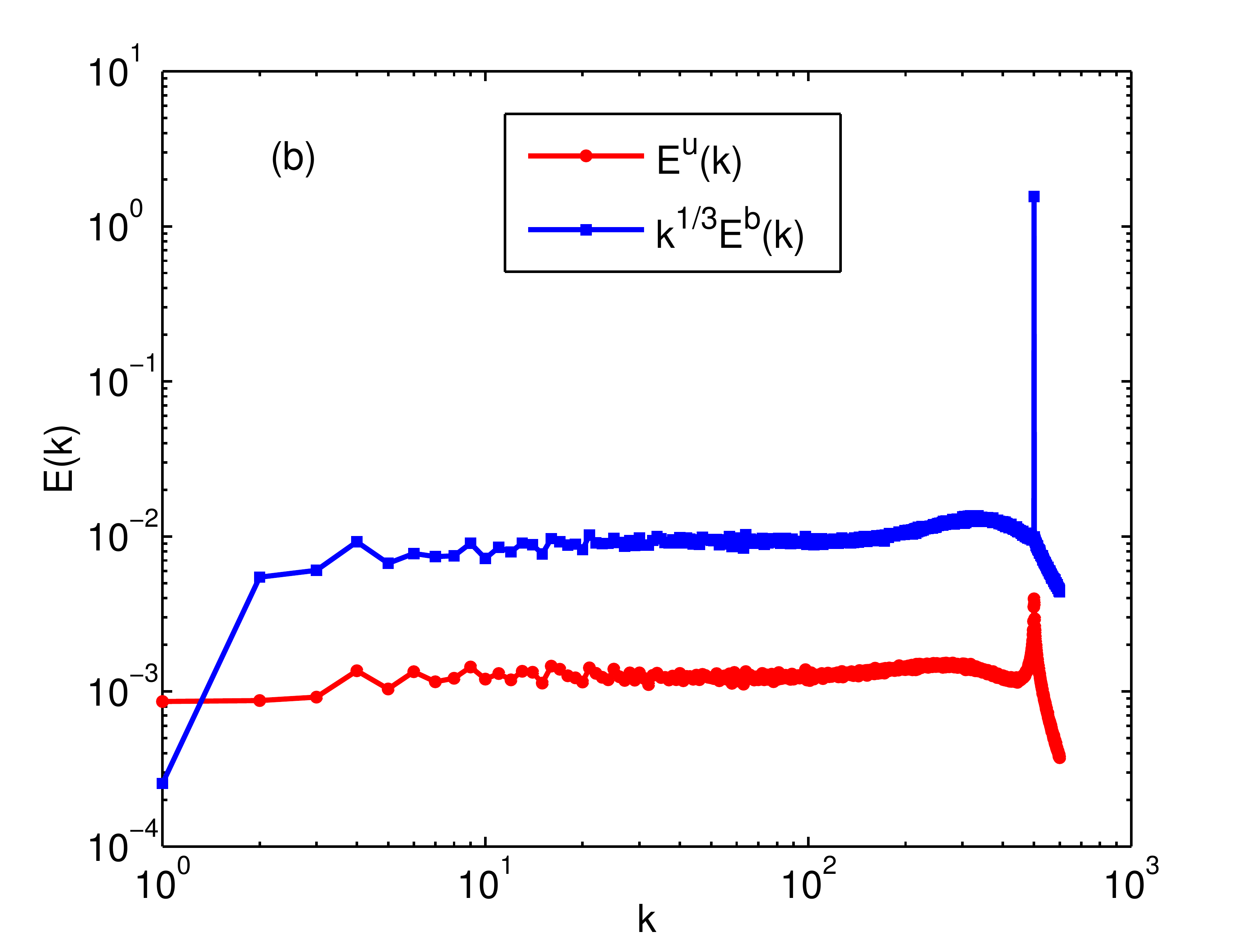} \\
\includegraphics[width=8cm,height=5cm]{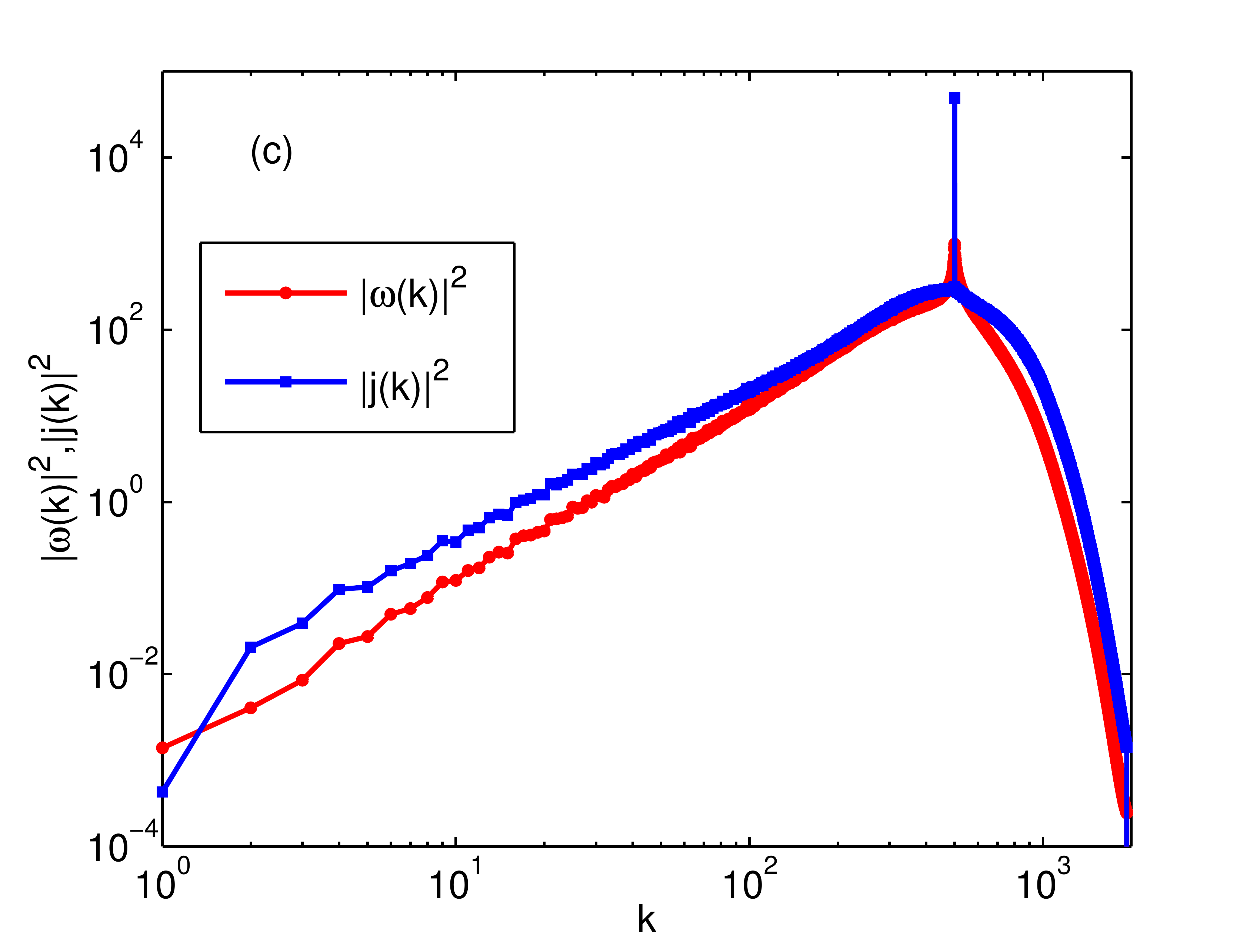}
\includegraphics[width=8cm,height=5cm]{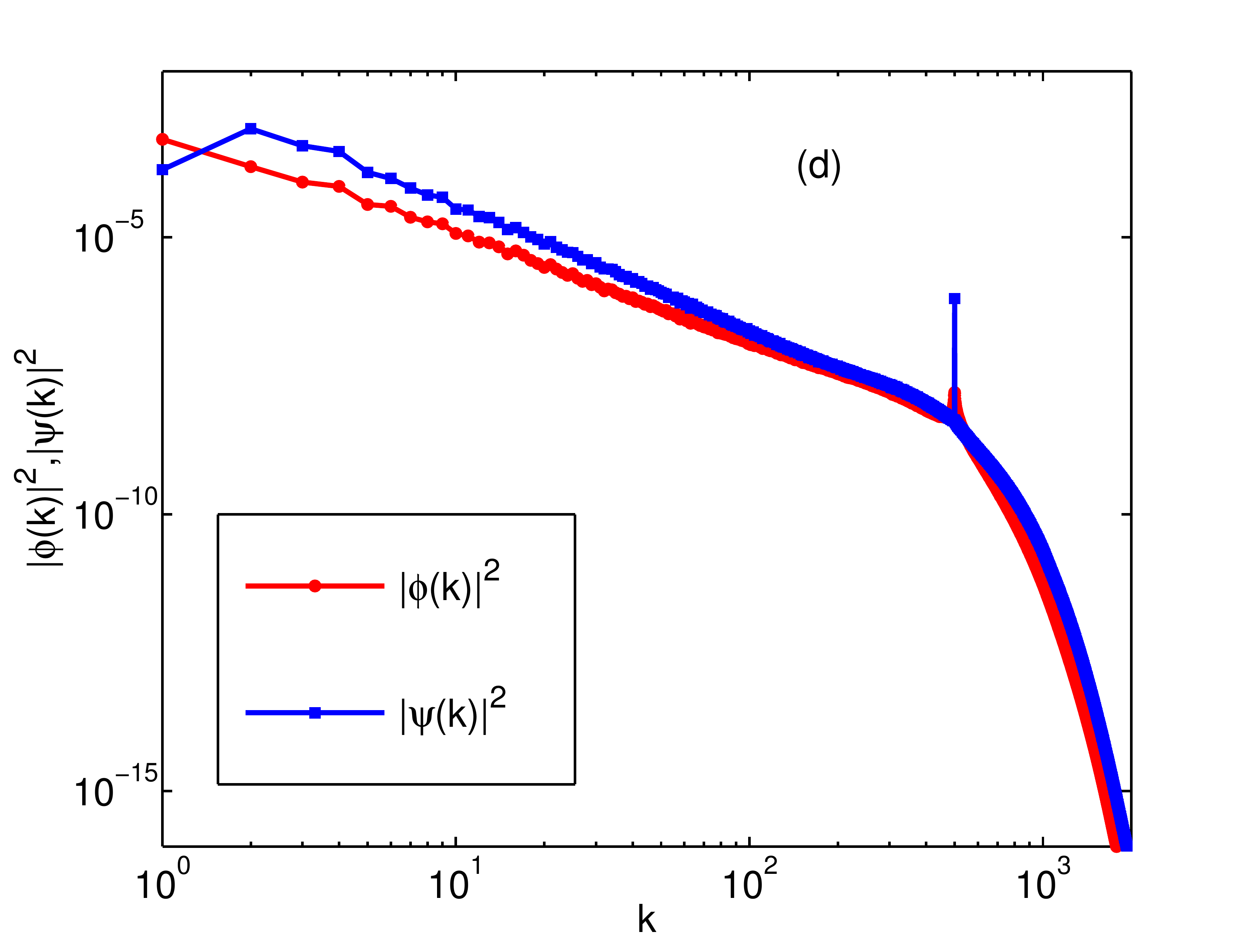}
\caption{(Color online) Log-log plots of spectra,for run R1 with
$k_{\rm inj} = 500$, versus the wave number $k$: (a)
kinetic-energy $E^u(k)$ (red curve) and magnetic-energy
$E^b(k)$ (blue curve); (b) the compensated spectra
$k^0E^u(k)$ (red curve) and $k^{1/3}E^b(k)$ (blue curve);
(c) vorticity $|\omega(k)|^2$ (red curve) and current
density $|j(k)|^2$ (blue curve) spectra; and (d) stream
function $|\phi(k)|^2$ (red curve) and magnetic potential
$|\psi(k)|^2$ (blue curve) spectra. These spectra are
averaged over thirty independent field configurations,
which are separated from one another by $\tau_{\rm eddy}$
(Table\ref{table:param}). A clear inverse-cascade inertial range
is visible in the interval $5 \le k \le 110$.}
\label{fig:spectra}
\end{figure*}

In Fig.~\ref{fig:spectra} we plot various spectra for run R1.
Figure~\ref{fig:spectra} (a) shows log-log plots of the kinetic-
and magnetic-energy spectra $E^u(k)$ (red curve) and $E^b(k)$
(blue curve), respectively, versus the wave number $k$, with
sharp peaks at the energy-injection wave number $k_{\rm inj} = 500$.
For $k < k_{\rm inj}$, these spectra display power-law regimes, which
are consistent with $E^u(k) \sim k^0$ and $E^b(k) \sim k^{-1/3}$,
as we can see from the compensated spectra of
Fig.~\ref{fig:spectra} (b), where the dashed line indicates the
range of $k$ (slightly more than a decade from $5 < k < 110$)
over which we obtain these power-law forms. These power laws are
cut off at very small values of $k$, because of the friction term
in Eq.(~\ref{eq:2dmhd}); and the hyperviscous term yields the mild
bottleneck maxima~\cite{bottle}, which are clearly
visible in the compensated spectra (Fig.~\ref{fig:spectra} (b))
close to and to the left of $k_{\rm inj}$.  Figure~\ref{fig:spectra}
(c) shows log-log plots of the vorticity (red curve) and current
(blue curve) spectra $|\omega(k)|^2$ and $|j(k)|^2$,
respectively, which follow simply from the energy spectra
mentioned above.  Plots of the stream-function and
magnetic-vector-potential spectra $|\phi(k)|^2$ (red curve) and
$|\psi(k)|^2$ (blue curve) are given in Fig.~\ref{fig:spectra}
(d). 

Figure~\ref{fig:spectra2} (a) is the analog, for run R2, of
Fig.~\ref{fig:spectra} (a), for run R1. A comparison of the
energy spectra in these two figures highlights important
differences between our runs R1 and R2. The energy spectra for run
R1 do not increase indefinitely as $k \to 0$, because of the
friction term, whereas those for run R2 do increase, because this
run does not use friction. The compensated spectra of
Fig.~\ref{fig:spectra} (b) (run R1) show bottleneck maxima, but
those of Fig.\ref{fig:spectra2} (b) (run R2) do not, because run
R1 uses hyperviscosity, but run R2 does
not~\cite{bottle}. 

\begin{figure}
\includegraphics[width=0.8\columnwidth]{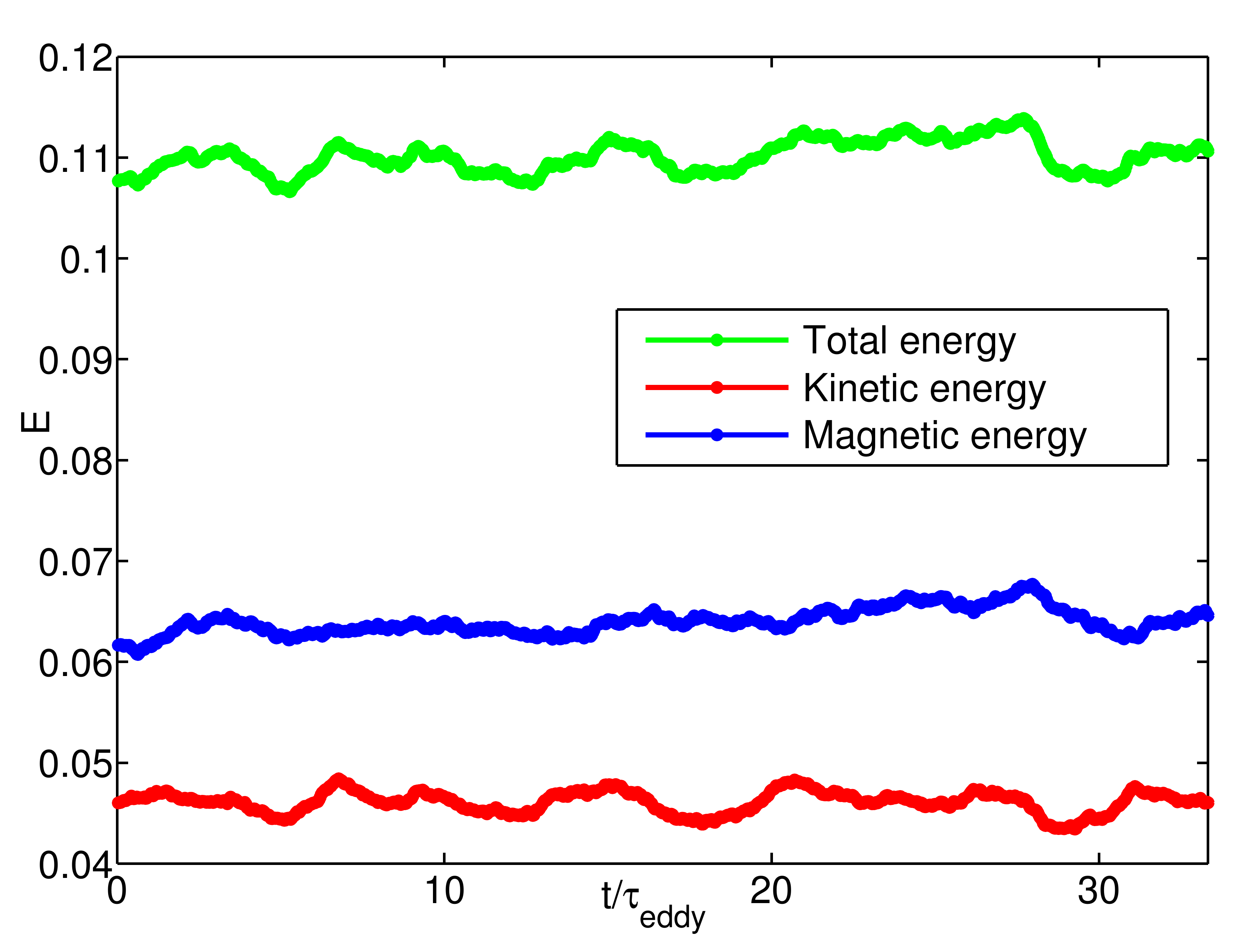}
\caption{(Color online) Plots, for run R2, of the total energy (green curve), 
magnetic-energy (blue curve), and kinetic-energy (red curve) versus time,
(measured in units of $\tau_{\rm eddy}$ (Table~\ref{table:param})).}
\label{fig:steady}
\end{figure}

\begin{figure*}[htbp]
\includegraphics[width=8.0cm,height=5.0cm]{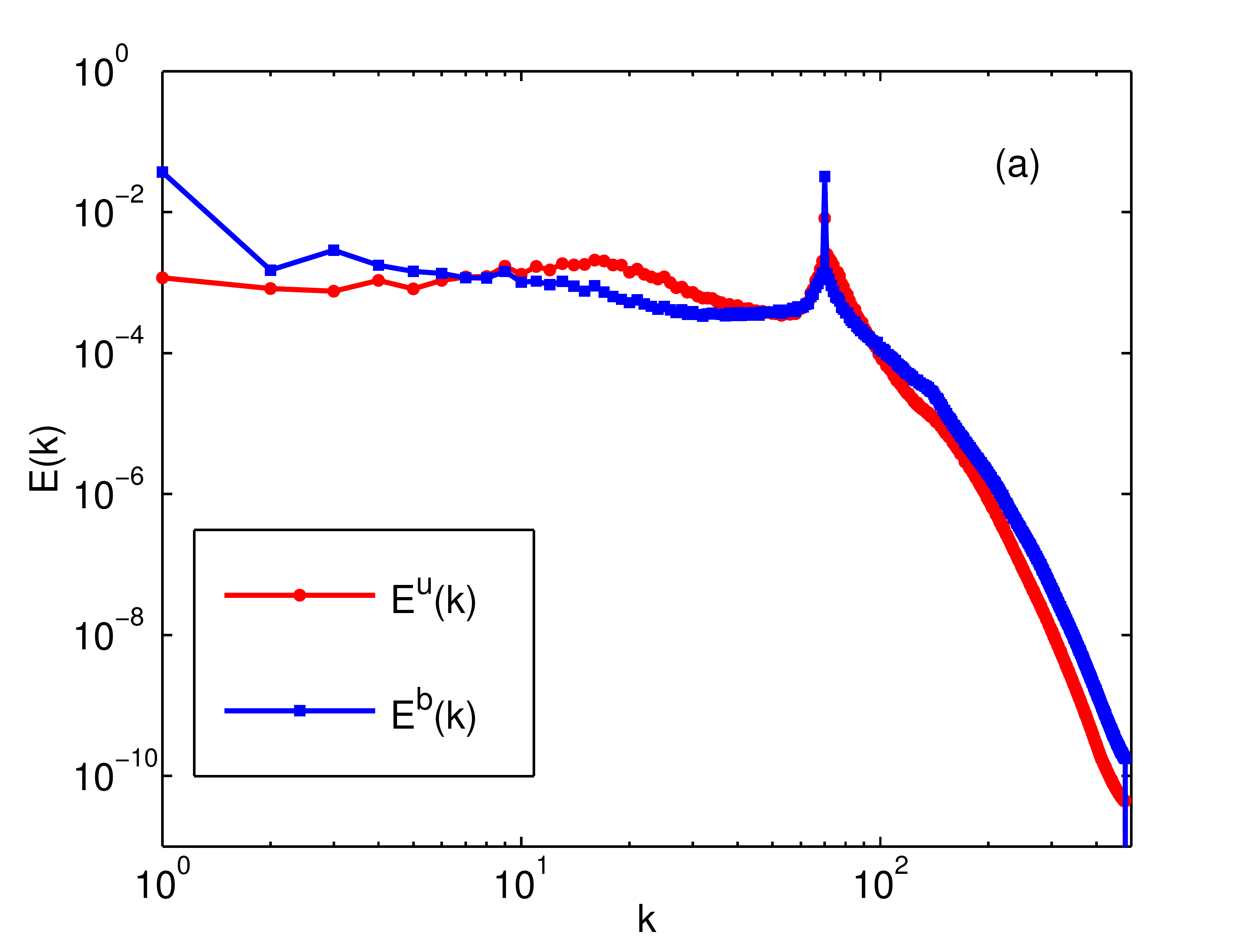}
\includegraphics[width=8.0cm,height=5.0cm]{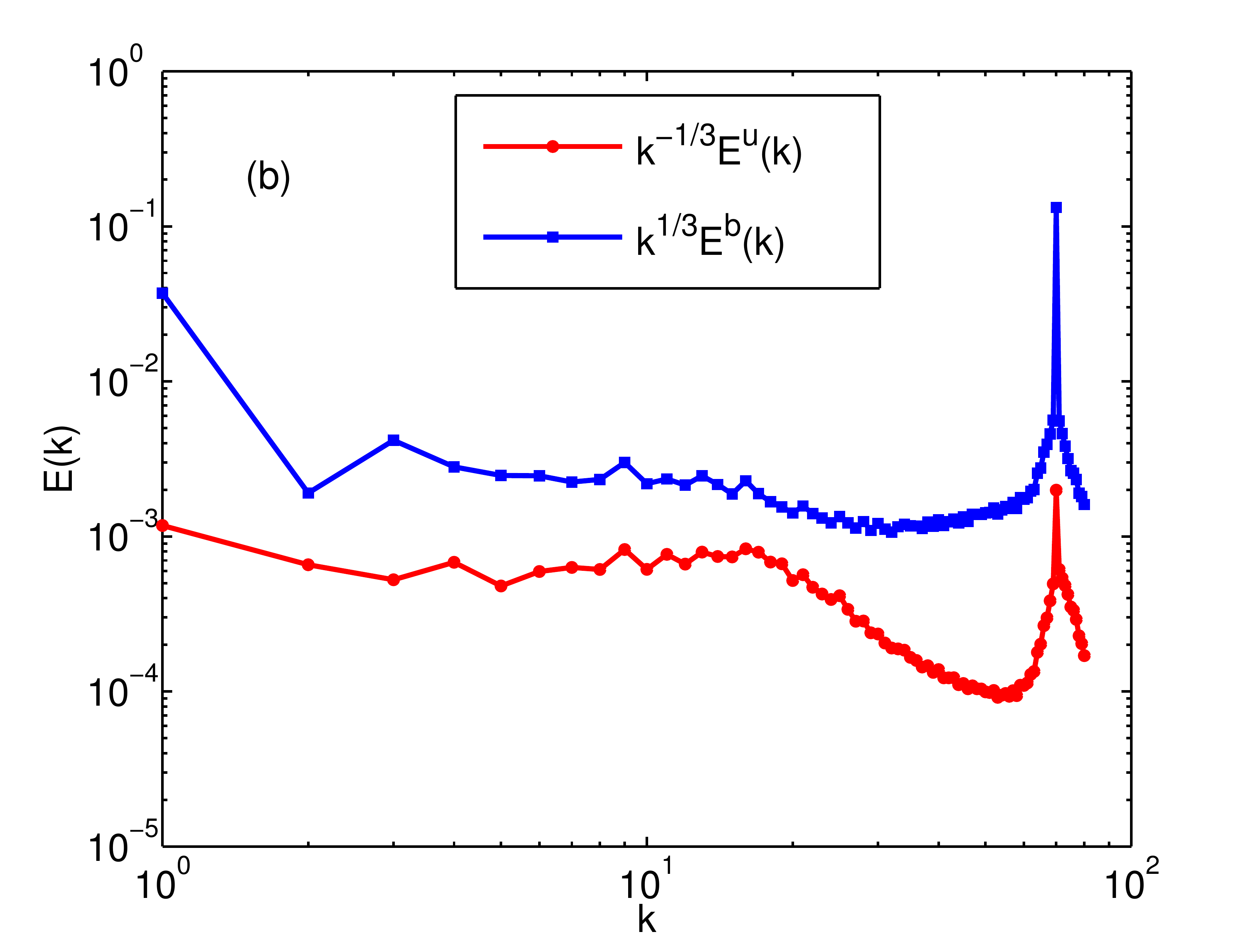} \\
\includegraphics[width=8.0cm,height=5.0cm]{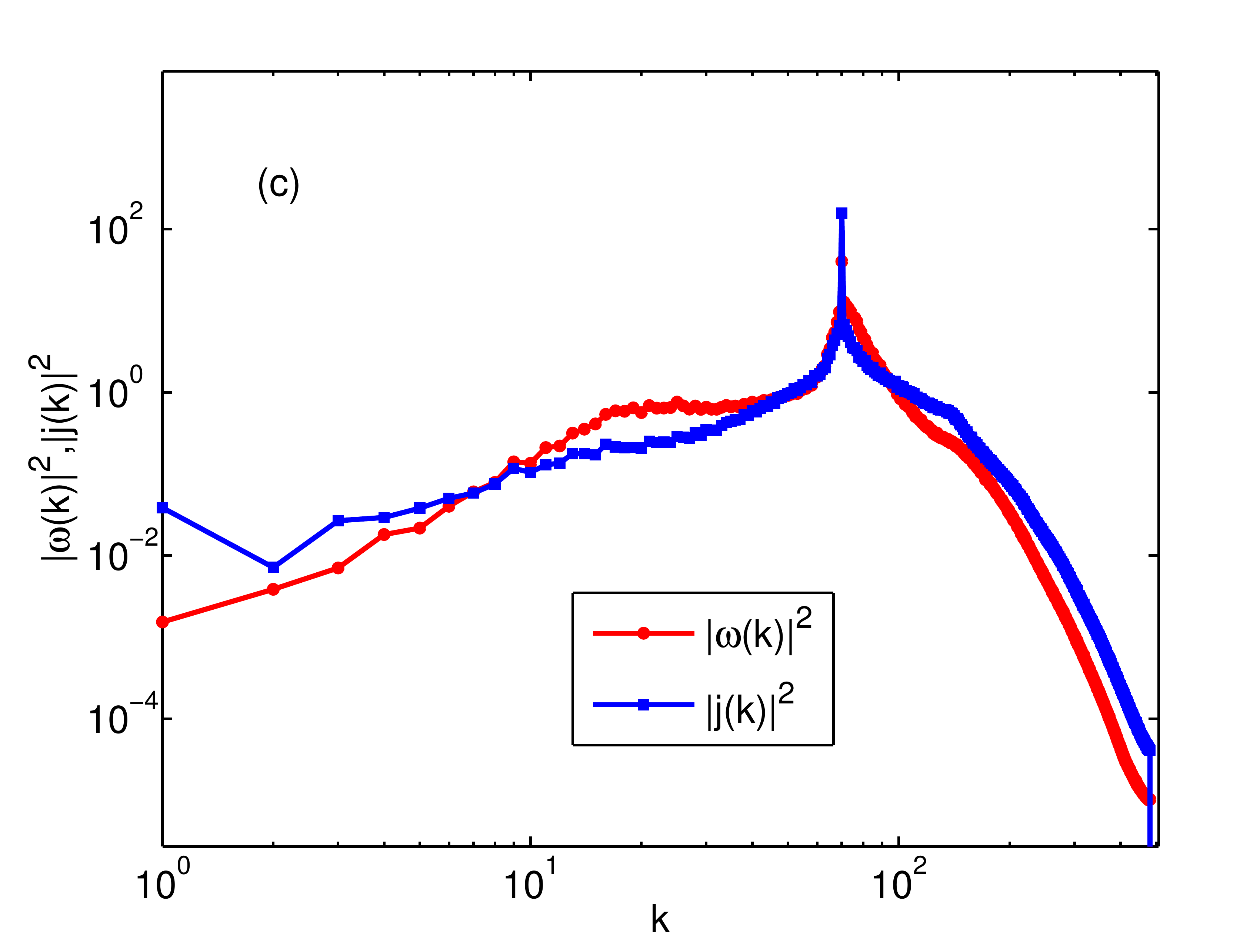}
\includegraphics[width=8.0cm,height=5.0cm]{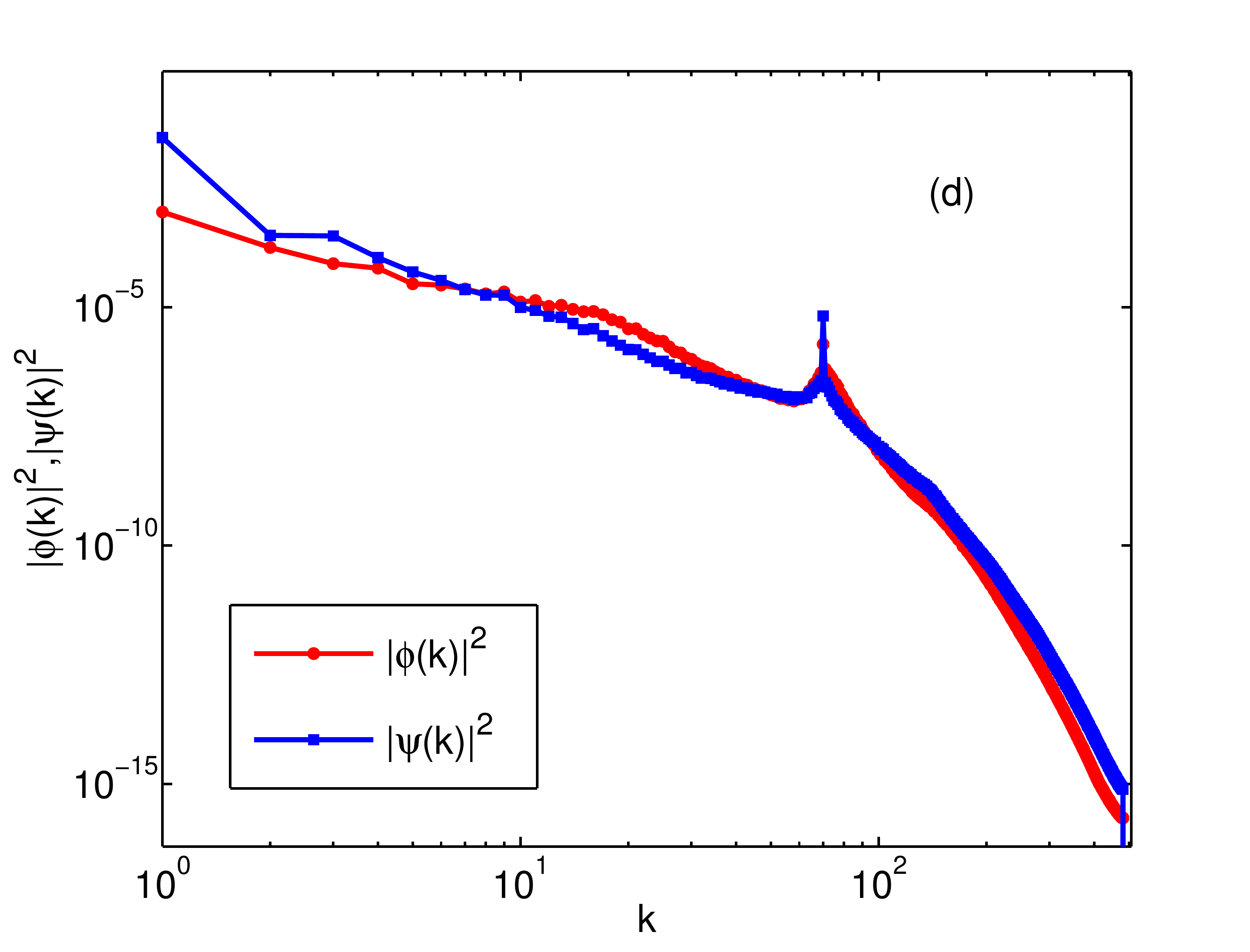}
\caption{(Color online) Log-log plots of spectra, for run R2,
versus the wave number $k$:  (a) kinetic-energy $E^u(k)$
(red curve) and magnetic-energy $E^b(k)$ (blue curve);
(b) the compensated spectra $k^{-1/3}E^u(k)$ (red curve)
and $k^{1/3}E^b(k)$ (blue curve); (c) the vorticity
spectrum $|\omega(k)|^2$ (red curve) and the current
density specturm $|j(k)|^2$ (blue curve); and (d) the
stream function spectrum $|\phi(k)|^2$ (red curve) and magnetic
potential spectrum $|\psi(k)|^2$ (blue curve).}
\label{fig:spectra2}
\end{figure*}

Another important difference between runs R1 and R2 is that the
latter takes a much longer time to reach a statistically steady
state than the former. \lbrack It has been noted in the context of 2D
fluid turbulence~\cite{chan12} that a simulation time of order
$1/(\nu k_1^2)$ is required to obtain such a steady state if
there is no friction (here $k_1$ is the smallest wave number in
the DNS).\rbrack ~This is why the spatial resolution
(and, therefore, the extent of the power-law range in the spectrum)
of run R2 is lower than that of run R1. In Fig.~\ref{fig:steady} we plot the total, kinetic,
and magnetic energies versus time, for run R2, and the last
$\simeq 33 \tau_{\rm eddy}$, to show that we have, indeed, reached a
statistically steady state (the total length of this DNS is
$\simeq 60 \tau_{\rm eddy}$). From
Figs.~\ref{fig:spectra2} (a) and (b) we see that the power-law
regimes in these kinetic- and magnetic-energy spectra have
scaling forms that are consistent with $E^u(k) \sim k^{1/3}$ and
$E^b(k) \sim k^{-1/3}$.  In Figs.~\ref{fig:spectra2} (d) we plot
the vorticity spectrum $|\omega(k)|^2$ and the current
density spectrum $|j(k)|^2$ , and in (e) we plot the
spectra of the stream function and magnetic
potential $|\phi(k)|^2$ and $|\psi(k)|^2$, respectively.

\subsection{One-point statistics}
\label{section_onepoint}

\begin{figure*}[htbp]
\includegraphics[width=8cm,height=5cm]{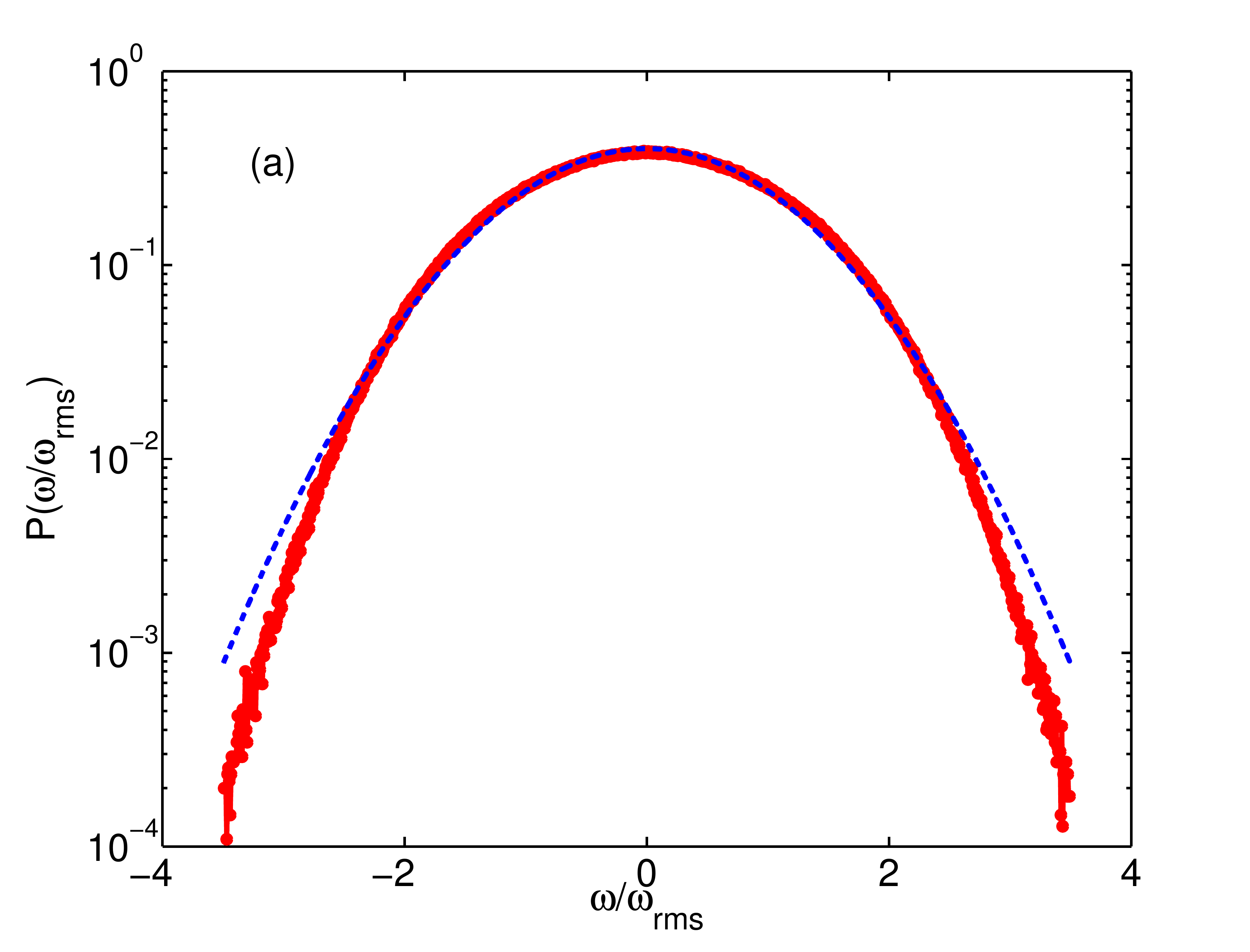}
\includegraphics[width=8cm,height=5cm]{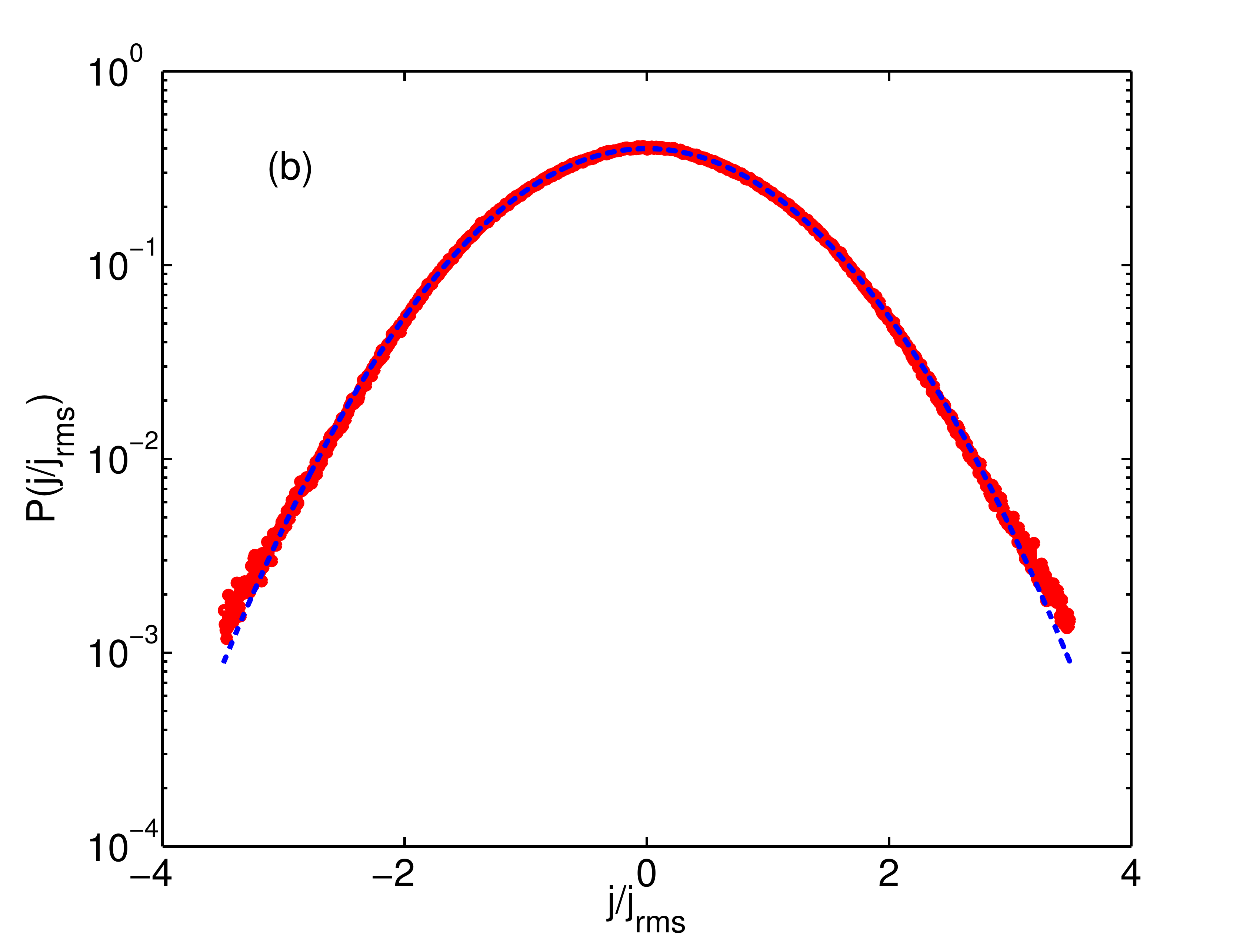} \\
\includegraphics[width=8cm,height=5cm]{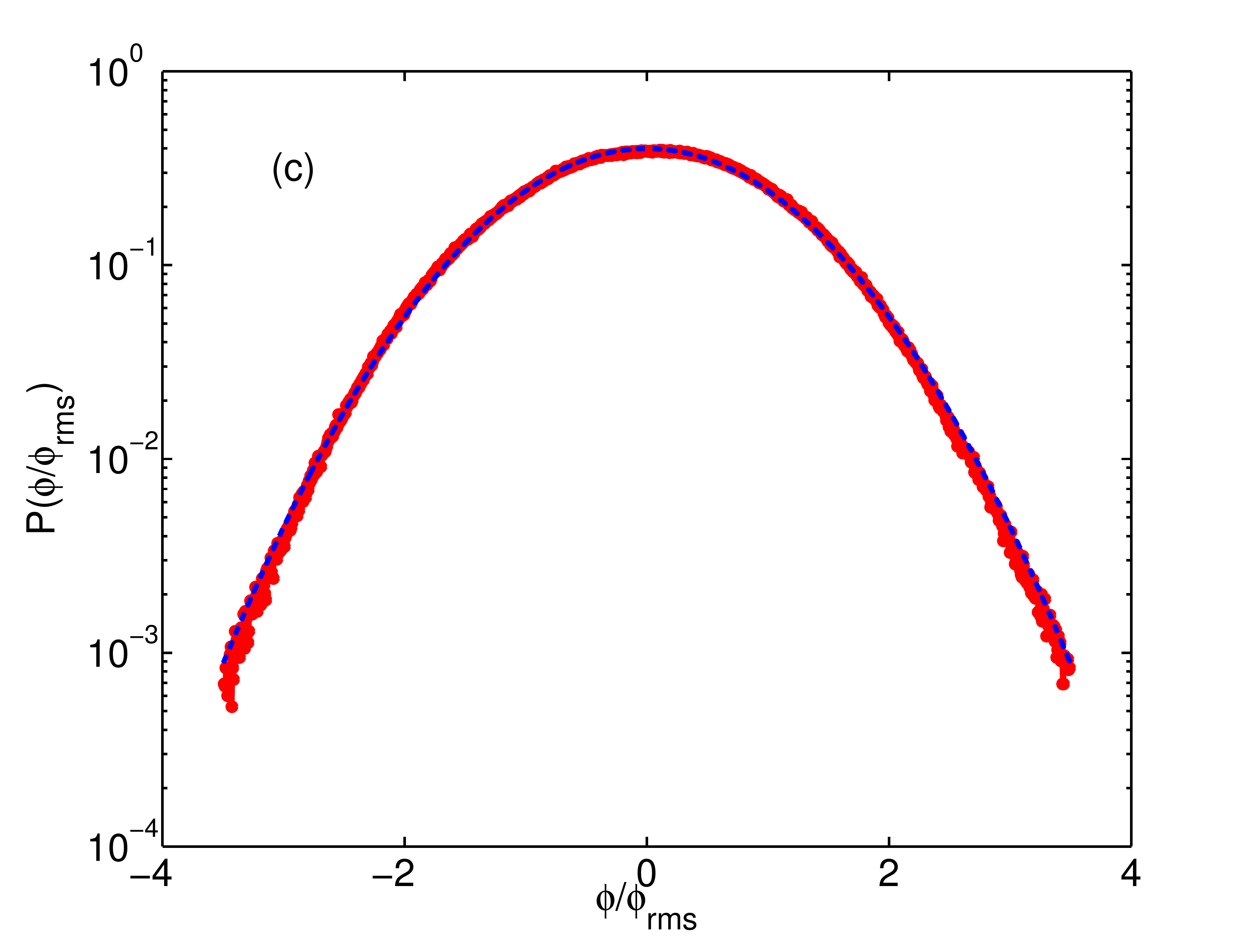}
\includegraphics[width=8cm,height=5cm]{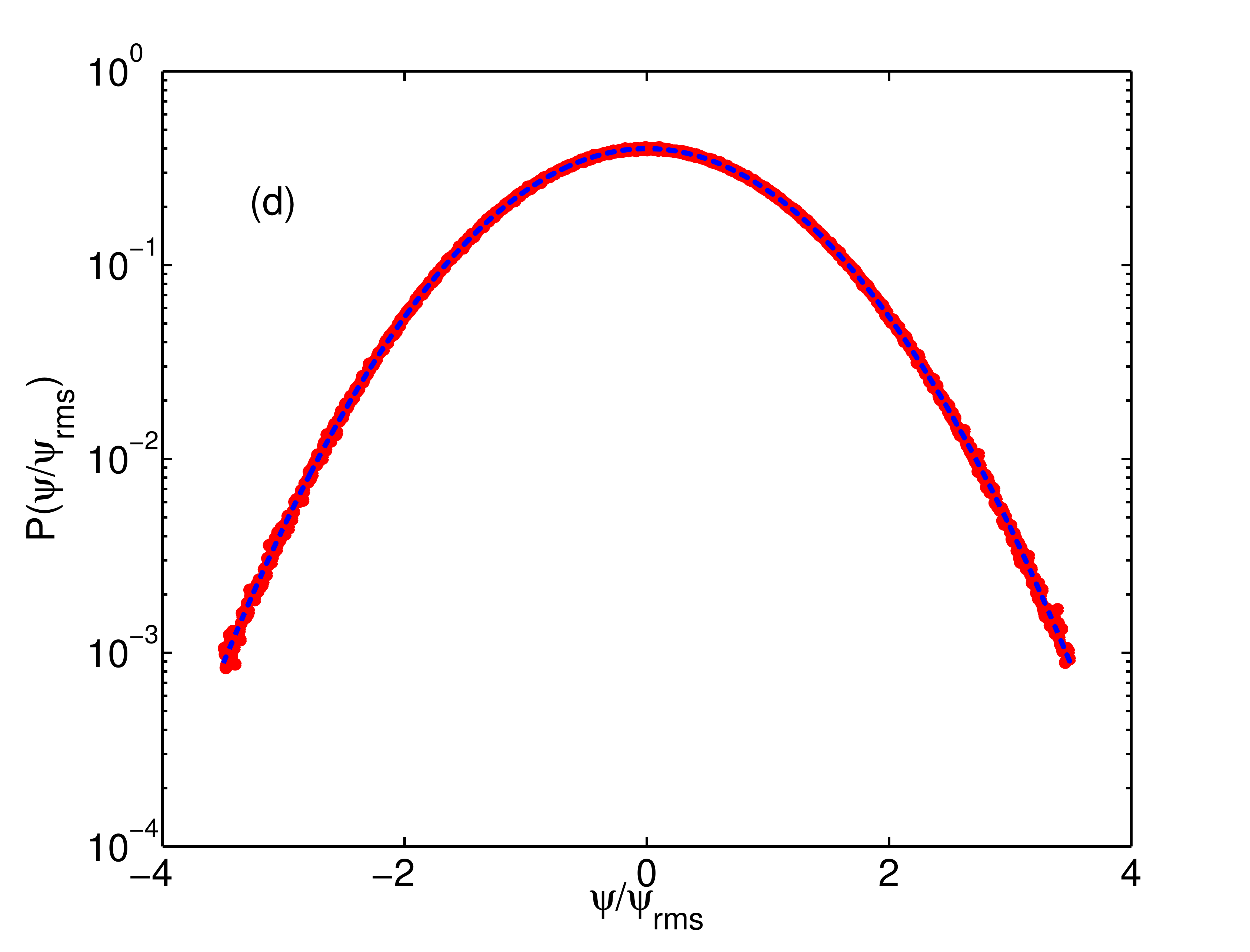}
\caption{(Color online) Semilog plots of the PDFs of (a) the
vorticity $\omega$, (b) the current density $j$, (c) the
stream function $\phi$, and (d) the magnetic potential
$\psi$. Blue dashed lines indicate Gaussian
distributions.  The deviations of our PDFs from Gaussian
ones are small and visible only in the tails of these
distributions (see text for their flatnesses).}
\label{fig:one_point_pdf}
\end{figure*}

In Figs.~\ref{fig:one_point_pdf} (a)-(d) we show, for Run R1,
PDFs (red curves) of $\omega, \, j,\, \phi,$ and $\psi$,
respectively, with Gaussian PDFs (blue dashed curves) for
comparison.  The PDFs of $\phi$ and $\psi$ do not show
significant deviations from Gaussian PDFs, but the tails of the
PDFs of $\omega$ and $j$ lie, respectively, above and below their
Gaussian counterparts. We confirm this by obtaining the kurtoses
or flatnesses $F_4(\omega) = \langle \omega^4 \rangle / \langle
\omega^2 \rangle^2$ and $F_4(j) = \langle j^4 \rangle / \langle
j^2 \rangle^2$; we find $F_4(\omega) = 2.7$ and $F_4(j) = 3.1$,
both of which differ significantly from the Gaussian (superscript
$G$) value $F_4^G = 3$. We obtain similar results for run R2. 

\begin{figure*}[htbp]
\includegraphics[width=8cm,height=5cm]{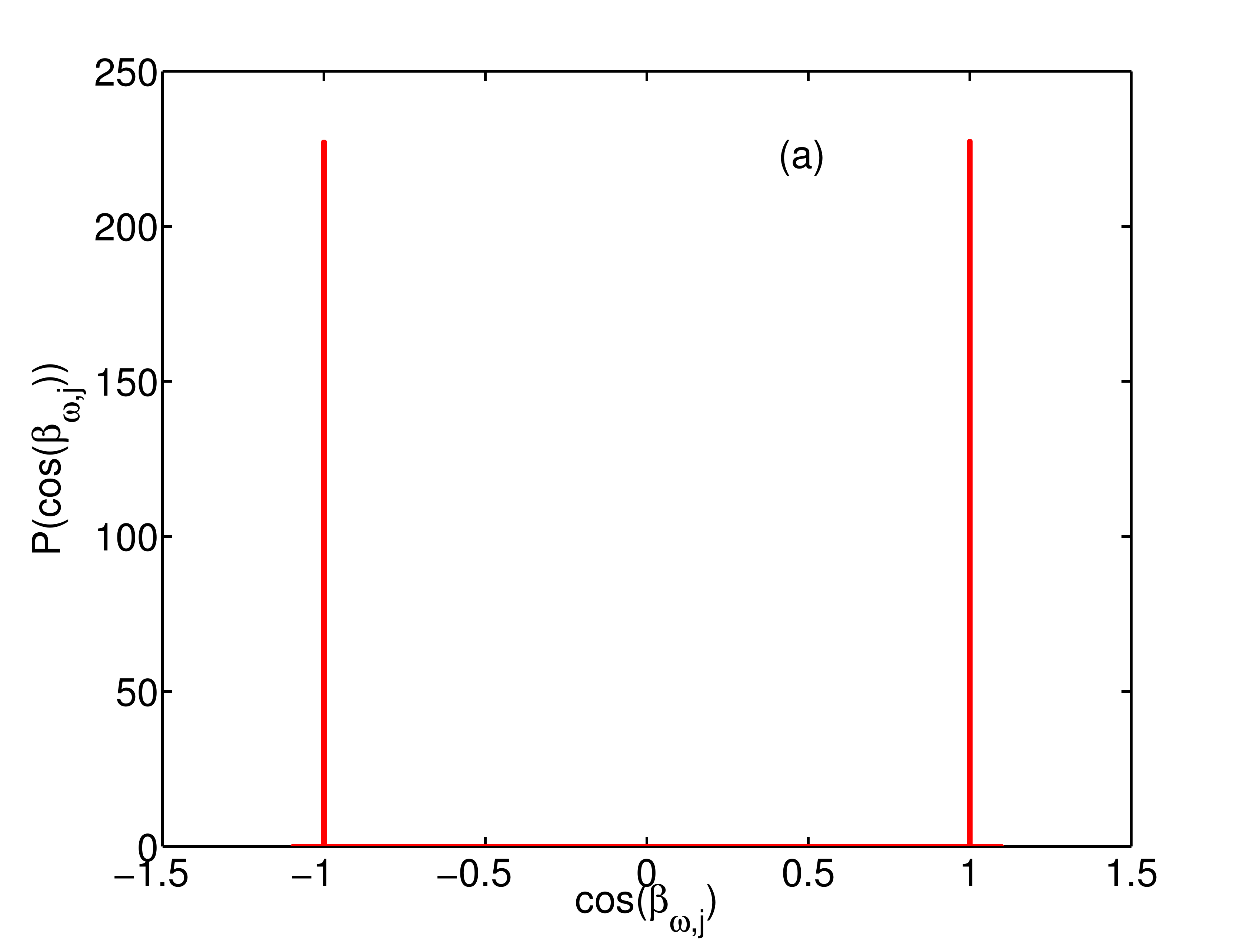}
\includegraphics[width=8cm,height=5cm]{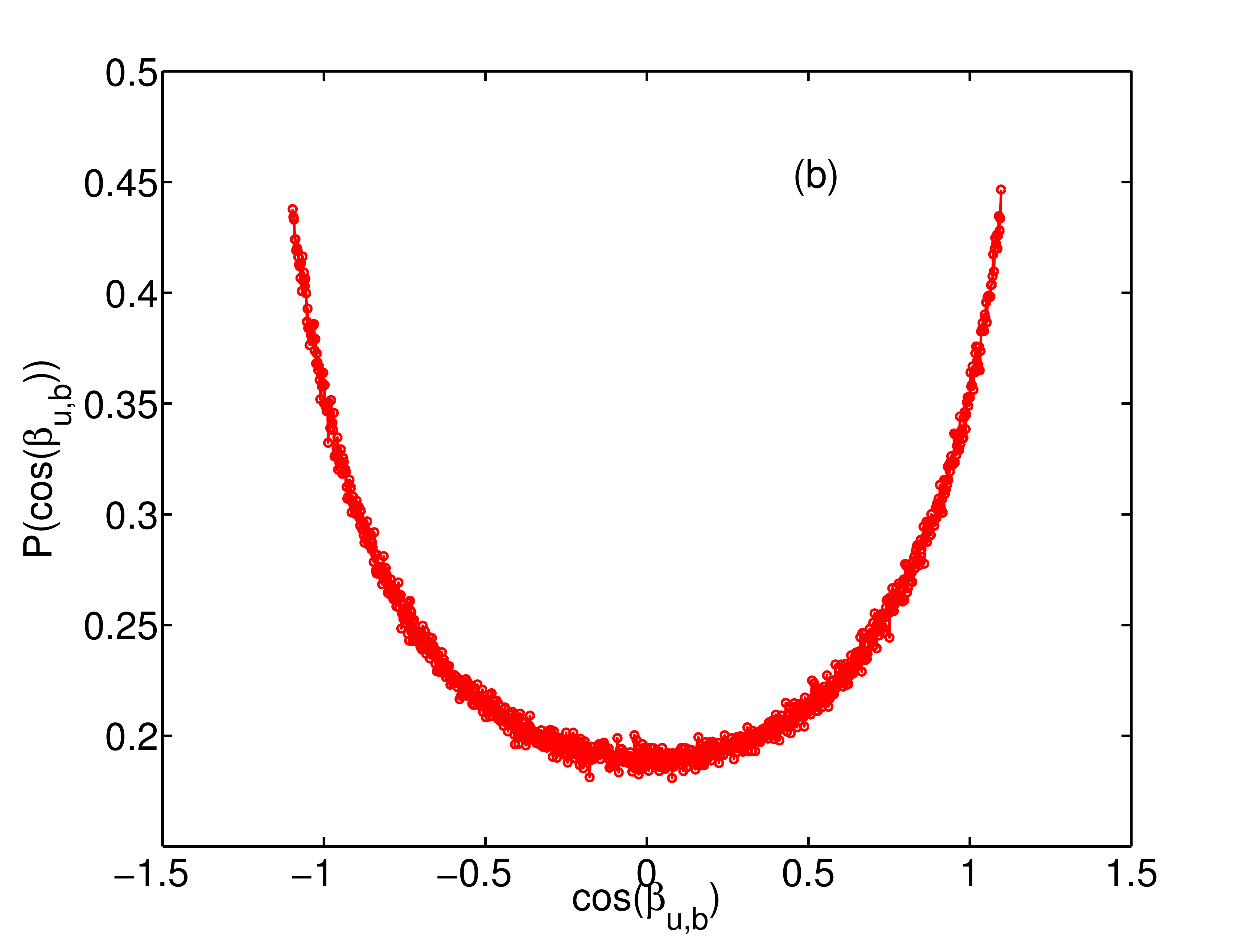}
\caption{(Color online) Plots of the PDFs for (a)
$\cos(\beta_{\omega,j})$, with $\beta_{\omega,j}$ the 
angle between $\omega$ and $j$ 
and $\cos(\beta_{\omega,j}) = \frac{\omega \cdot j}{|\omega||j|}$, and
(b) $\cos(\beta_{u,b})$, with $\beta_{u,b}$ the 
angle between ${\bf u}$ and ${\bf b}$, i.e., 
$\cos(\beta_{u,b}) = \frac{{\bf u}\cdot {\bf b}}{|{\bf u}||{\bf b}|}$.}
\label{fig:alignment}
\end{figure*}

In Figs.~\ref{fig:alignment} (a) and (b) we present, for run R1,
PDFs of the cosines of the angles $\beta_{\omega,j}$ and
$\beta_{u,b}$ between (a) $\omega$ and $j$ and (b)  $\bf u$ and
$\bf b$, respectively. These PDFs quantify the degree of alignment between
these vectors. From  Fig.~\ref{fig:alignment} (a) we see that
$\omega$ and $j$, which are perpendicular to the simulation
domain and can be viewed as pseudoscalars, are either perfectly
aligned or antiparallel.  The PDF of Fig.~\ref{fig:alignment}
(b) shows that $\bf u$ and $\bf b$, which lie in the 2D
simulation plane, have a greater tendency to be aligned or
antiparallel than to be orthogonal; and the PDF of
$\cos(\beta_{u,b})$ is symmetrical about its minimum at $\beta_{u,b}=0$. We obtain similar
results for run R2. For earlier studies of such alignment PDFs we
refer the reader to Ref.~\cite{alignment}, which invertigates 
field alignments for decaying 3D and 2D MHD turbulence.

\begin{figure*}[htbp]
\includegraphics[width=1.0\columnwidth]{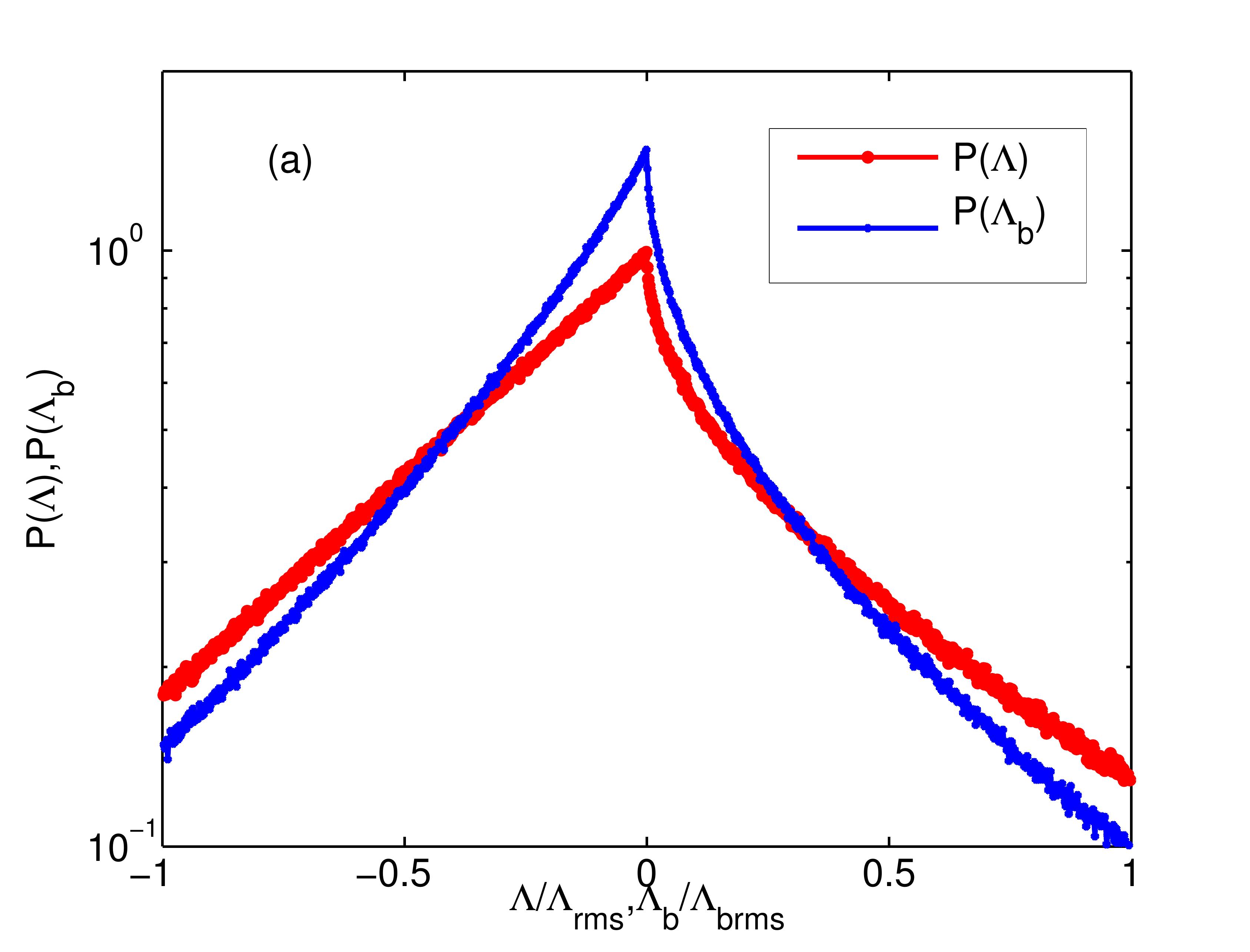}
\includegraphics[width=1.0\columnwidth]{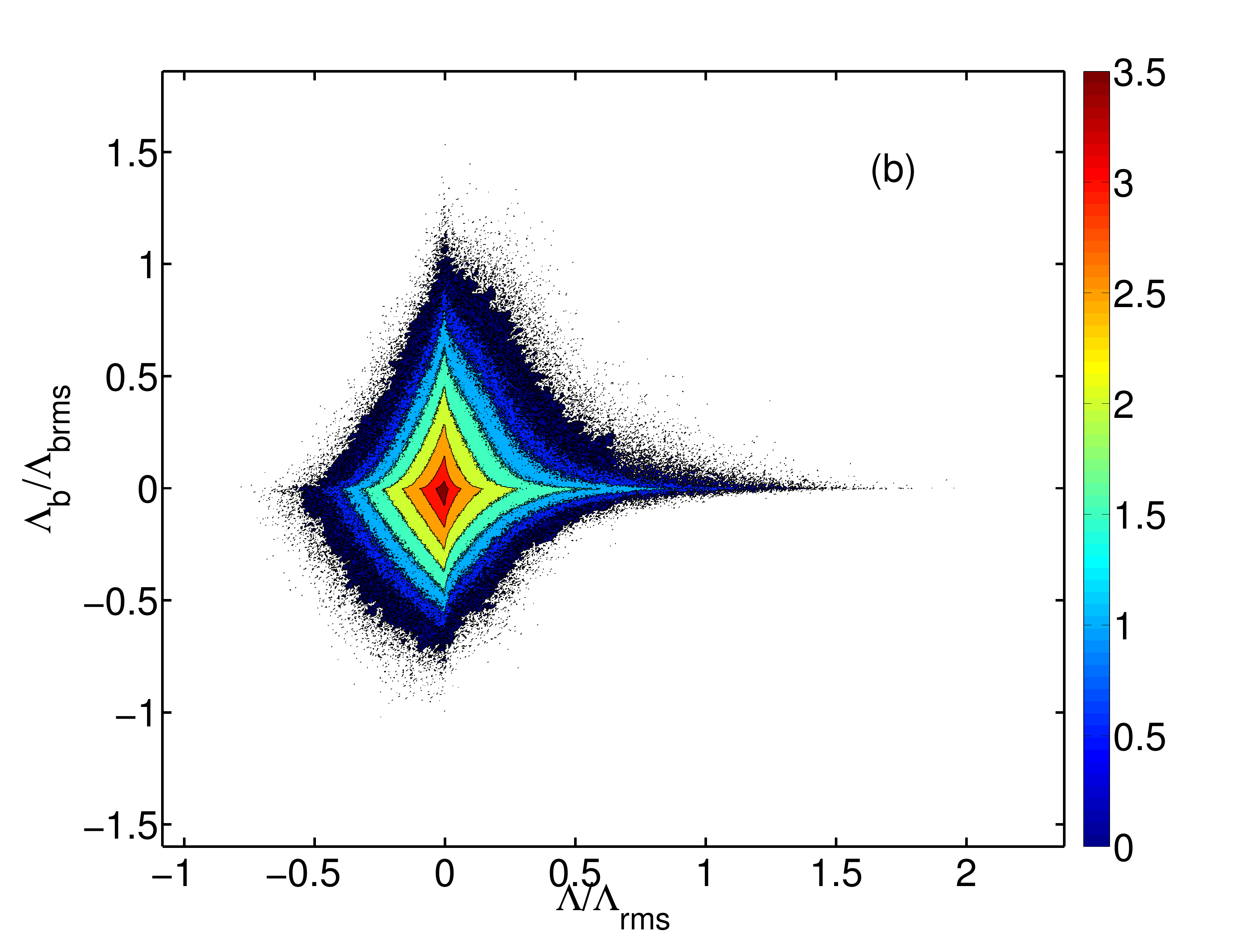}
\caption{(Color online) (a) The PDF of $\Lambda$ (red curve) and
$\Lambda_b$ (blue curve), and (b) the joint PDF of $\Lambda$ and
$\Lambda_b$, for run R1, with the axes rescaled by the rms value of the
quantity; the colorbar is logarithmic.}
\label{fig:lambda1}
\end{figure*}

\begin{figure*}[htbp]
\includegraphics[width=1.0\columnwidth]{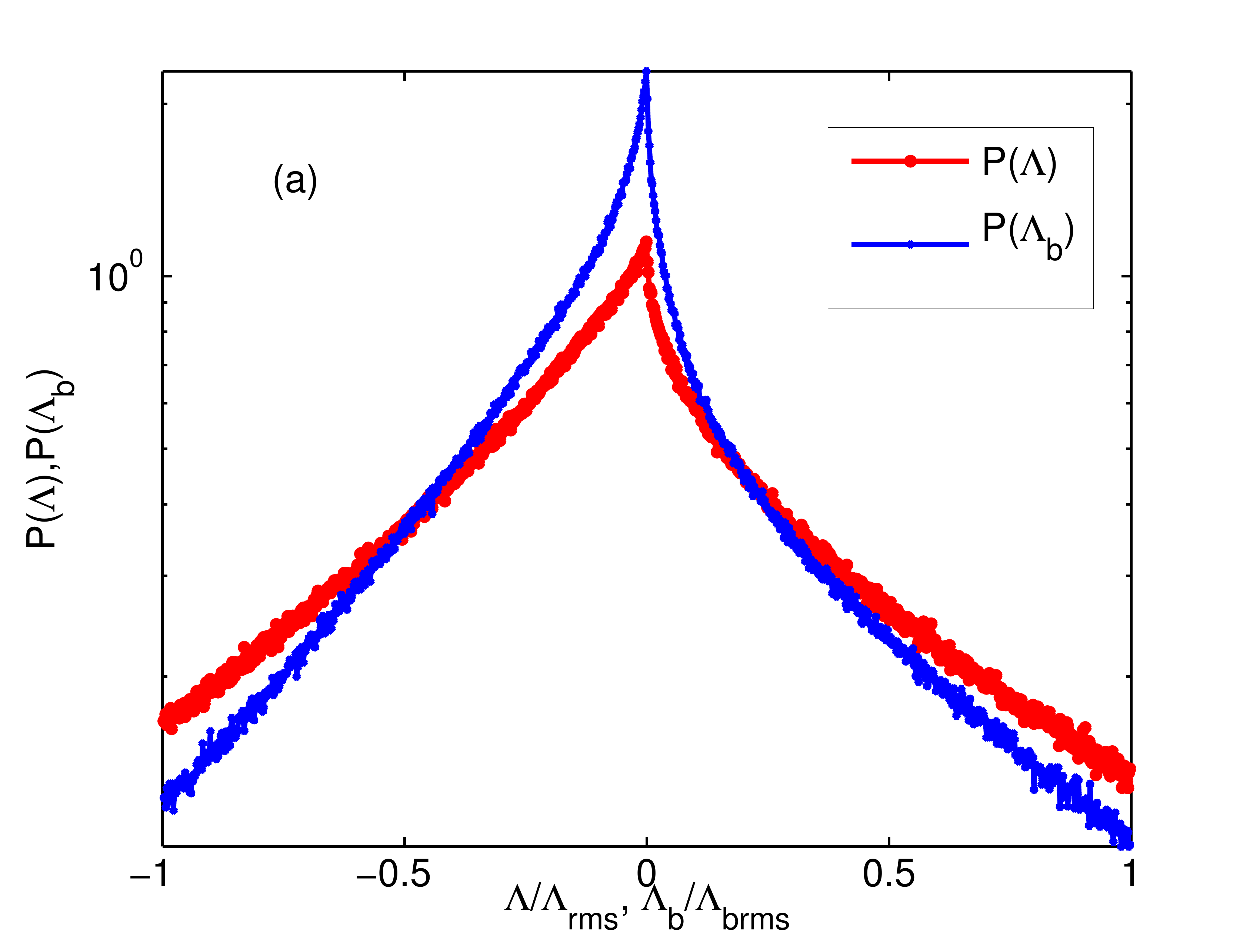}
\includegraphics[width=1.0\columnwidth]{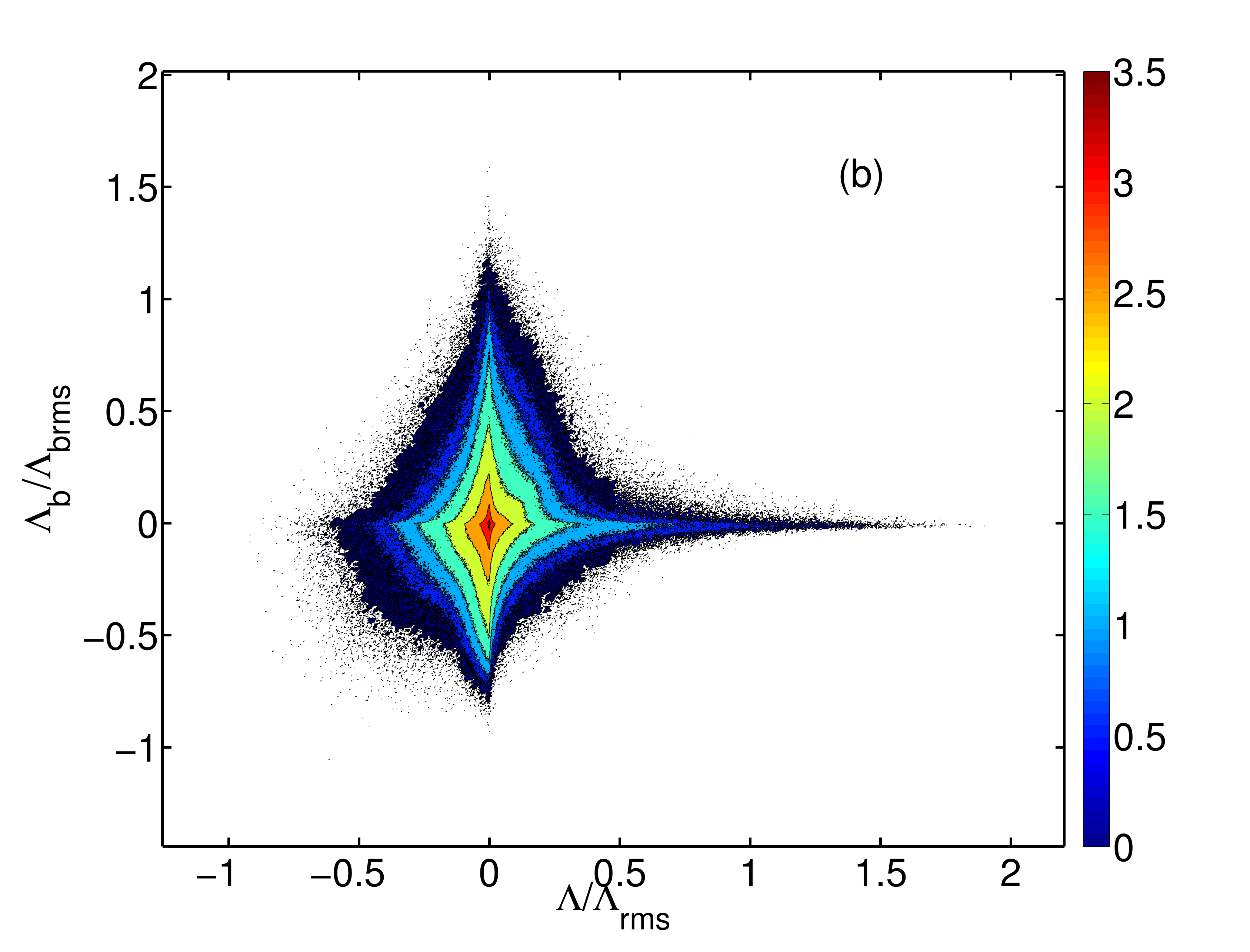}
\caption{(Color online) (a) The PDF of $\Lambda$ (red curve) and
$\Lambda_b$ (blue curve), and (b) the joint PDF of $\Lambda$ and
$\Lambda_b$, for run R2, with the axes rescaled by the rms value of the
quantity; the colorbar is logarithmic.}
\label{fig:lambda2}
\end{figure*}

We calculate $\Lambda$ and $\Lambda_b$ by using Eqs.
(\ref{eq:lamb1}) and (\ref{eq:lamb2}) and then obtain their
PDFs  ${\mathcal P}(\Lambda)$ and ${\mathcal P}(\Lambda_b)$,
which are shown, respectively, by red and blue curves in
Fig.~\ref{fig:lambda1} (a) (for run R1) and
Fig.~\ref{fig:lambda2} (a) (for run R2). There are two
qualitative similarities between ${\mathcal P}(\Lambda)$ and
${\mathcal P}(\Lambda_b)$ for both runs R1 and R2: all these PDFs
show a cusp at $\Lambda = 0$ or $\Lambda_b = 0$; and all of them
have tails that can be fit to exponentials over the range of
values we consider. For runs R1 (subscript $1$) and R2
(subscript $2$)  we parametrize the left (subscript $l$)  and
right (subscript $r$) tails of these PDFs as follows:
\begin{eqnarray} 
{\mathcal P}(\Lambda) &\sim& \exp(-\xi^u_{1,l}\Lambda);\, 
{\mathcal P}(\Lambda) \sim \exp(-\xi^u_{1,r}\Lambda); \nonumber \\ 
{\mathcal P}(\Lambda_b) &\sim& \exp(-\xi^b_{1,l}\Lambda_b);\,
{\mathcal P}(\Lambda_b) \sim \exp(-\xi^b_{1,l}\Lambda_b); \\ \nonumber
\end{eqnarray} 
here, the superscripts $u$ and $b$ denote velocity and magnetic
fields. We find $\xi^u_{1,l} \simeq 1.7$, $\xi^u_{1,r} \simeq
1.5$, $\xi^b_{1,l} \simeq 2.0$, and $\xi^b_{1,r} \simeq 2.0$, and
$\xi^u_{2,l} \simeq 1.6$, $\xi^u_{2,r} \simeq 1.5$, $\xi^b_{2,l}
\simeq 2.2$, and $\xi^b_{2,r} \simeq 1.9$.

We show the joint PDF ${\mathcal P}(\Lambda,\Lambda_b)$ of
$\Lambda$ and $\Lambda_b$ in Figs.~\ref{fig:lambda1} (b) and
~\ref{fig:lambda2} (b), for runs R1 and R2, respectively, by
using filled contours (and a logarithmic color scale). The joint
PDF can be divided into the following four regions: (1) $\Lambda
> 0$ and $\Lambda_b > 0$, where our 2D MHD flows are dominated by
vorticity and current; (2) $\Lambda > 0$ and $\Lambda_b < 0$,
where the flow is dominated by vorticity and magnetic strain
rate; (3) $\Lambda < 0$ and $\Lambda_b > 0$, where the flow is
dominated by the fluid strain and the current; (4) $\Lambda < 0$
and $\Lambda_b < 0$, where the flow is dominated by fluid and
magnetic rates of strain. The joint PDF ${\mathcal
P}(\Lambda,\Lambda_b)$ for run R1 (Fig.~\ref{fig:lambda1} (b)) is
qualitatively similar to its counterpart for R2
(Fig.~\ref{fig:lambda2} (b)). These joint PDFs have sharp peaks at
$\Lambda = \Lambda_b = 0$ and ridges that seperate the four regions 
mentioned above. 

\begin{figure*}[htbp]
\includegraphics[width=1.0\columnwidth]{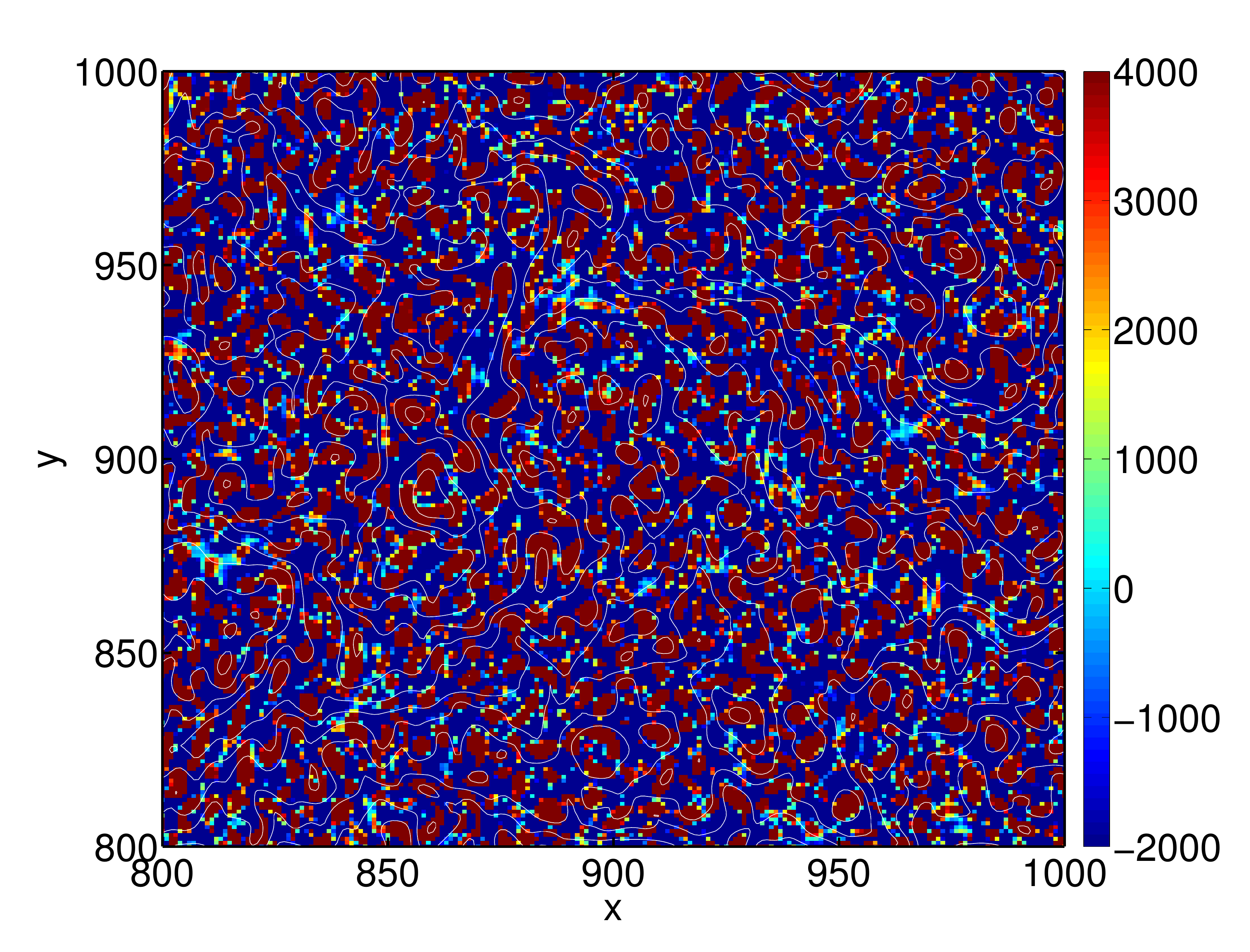}
\includegraphics[width=1.0\columnwidth]{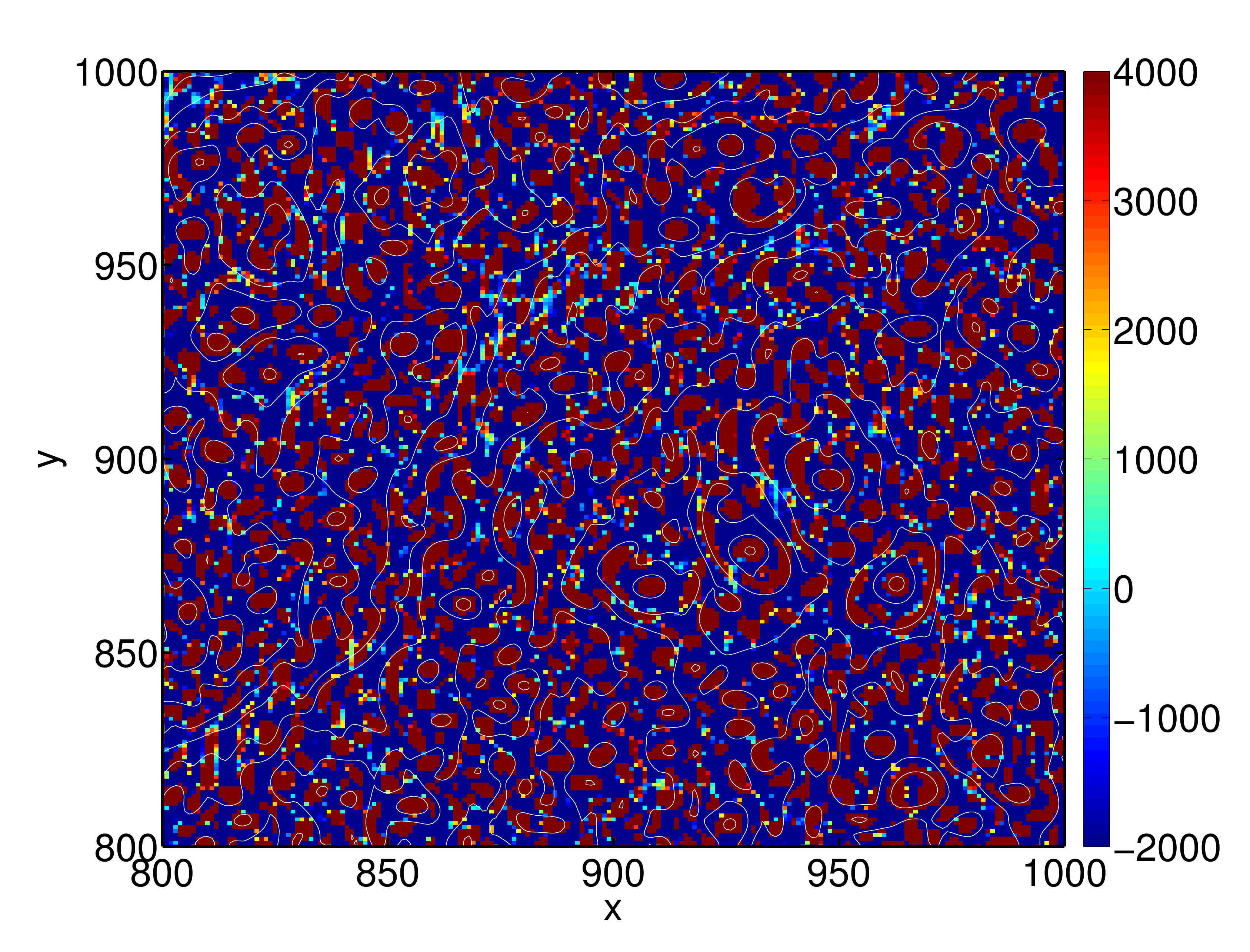}
\caption{(Color online) Left panel: Overlaid plot of the contours
of $\phi$ on a pseudocolor plot of $\Lambda$; right
panel: overlaid plot of the contours of $\psi$ on a
pseudocolor plot of $\Lambda_b$; these are for the run
R1.  The $\Lambda$ and $\Lambda_b$ fields have been
filtered in such a way that all values of
$\Lambda,\Lambda_b \ge 4000$ are set to
$\Lambda,\Lambda_b = 4000$ and $\Lambda,\Lambda_b \le
-2000$ are set to $\Lambda,\Lambda_b = -2000$; these
plots are for run R1.}
\label{fig:overlaid1}
\end{figure*}

\begin{figure*}[htbp]
\includegraphics[width=1.0\columnwidth]{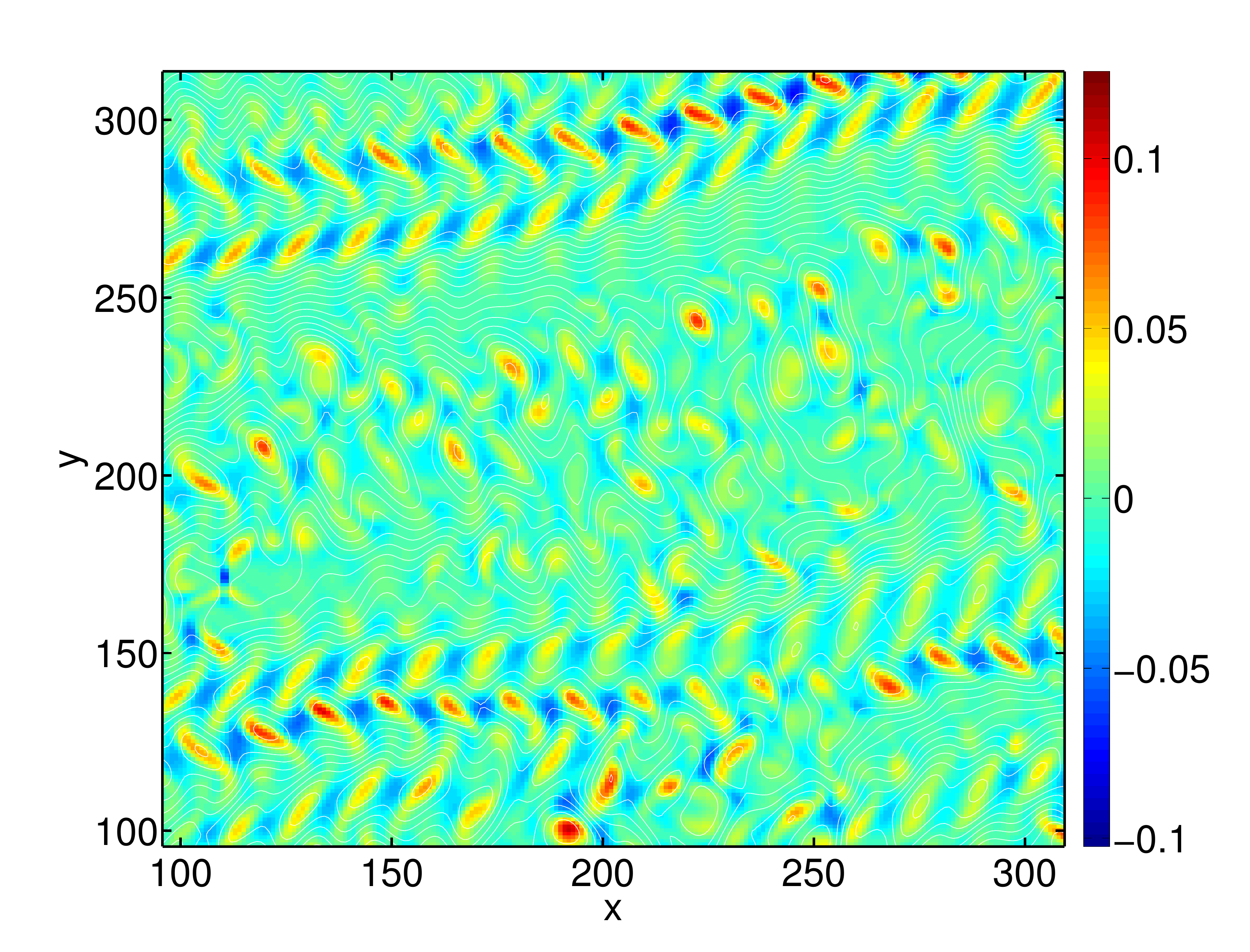}
\includegraphics[width=1.0\columnwidth]{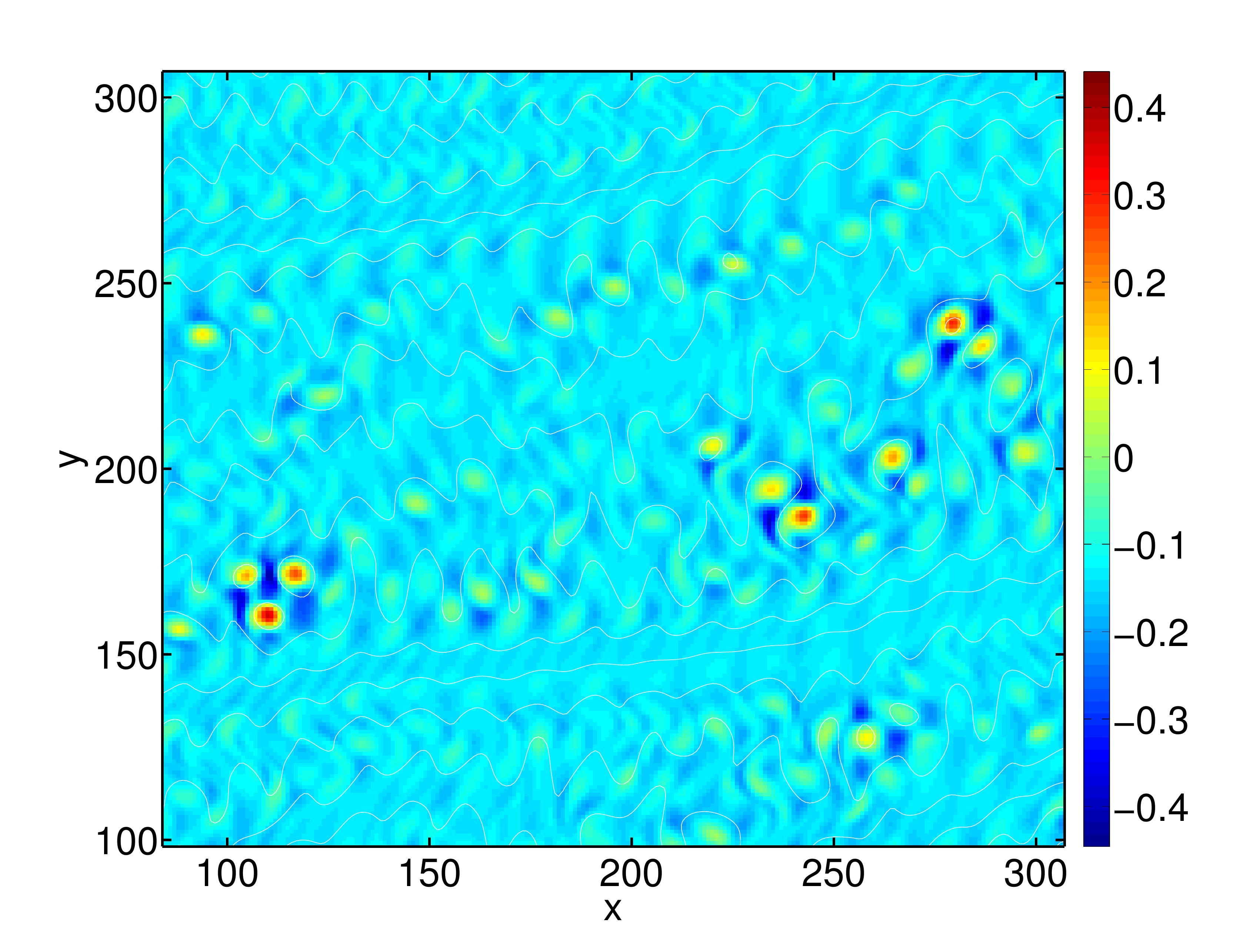}
\caption{(Color online) Left panel: Overlaid plot of the contours
of $\phi$ on a pseudocolor plot of $\Lambda$; right
panel: overlaid plot of the contours of $\psi$ on a
pseudocolor plot of $\Lambda_b$; these plots are for the run
R2.}
\label{fig:overlaid2}
\end{figure*}

\begin{figure*}[htbp]
\includegraphics[width=8cm,height=5cm]{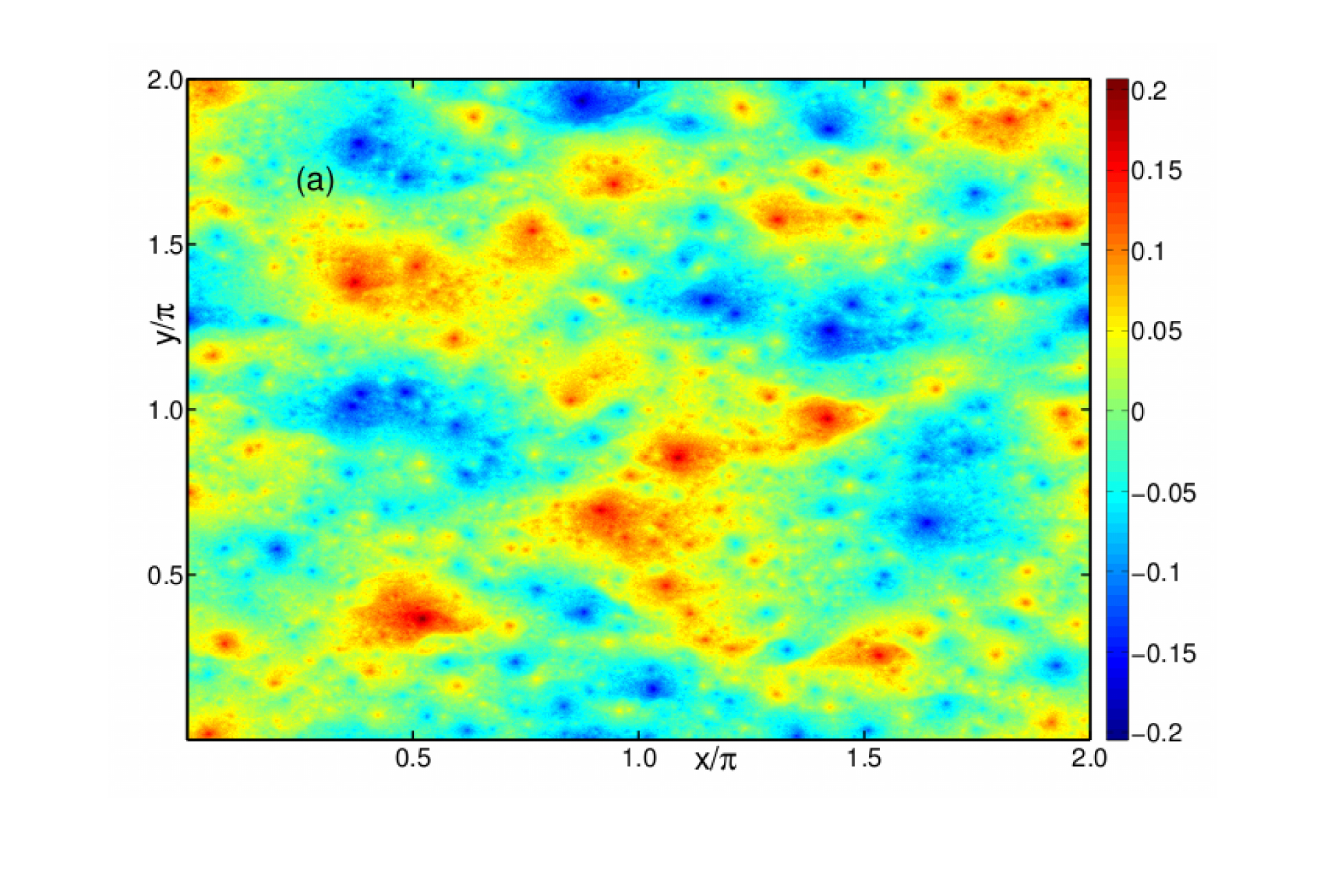}
\includegraphics[width=8cm,height=5cm]{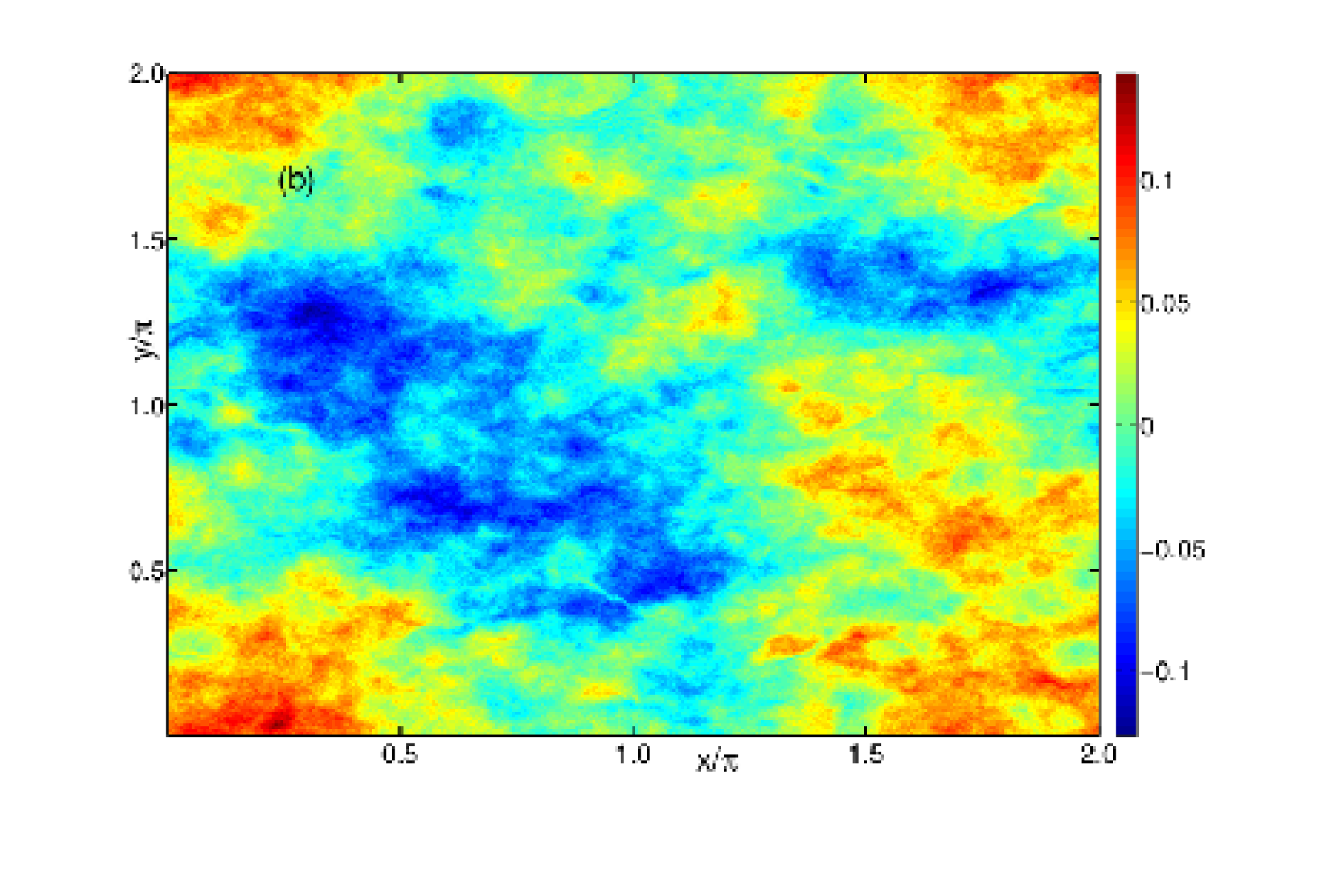}\\
\includegraphics[width=8cm,height=5cm]{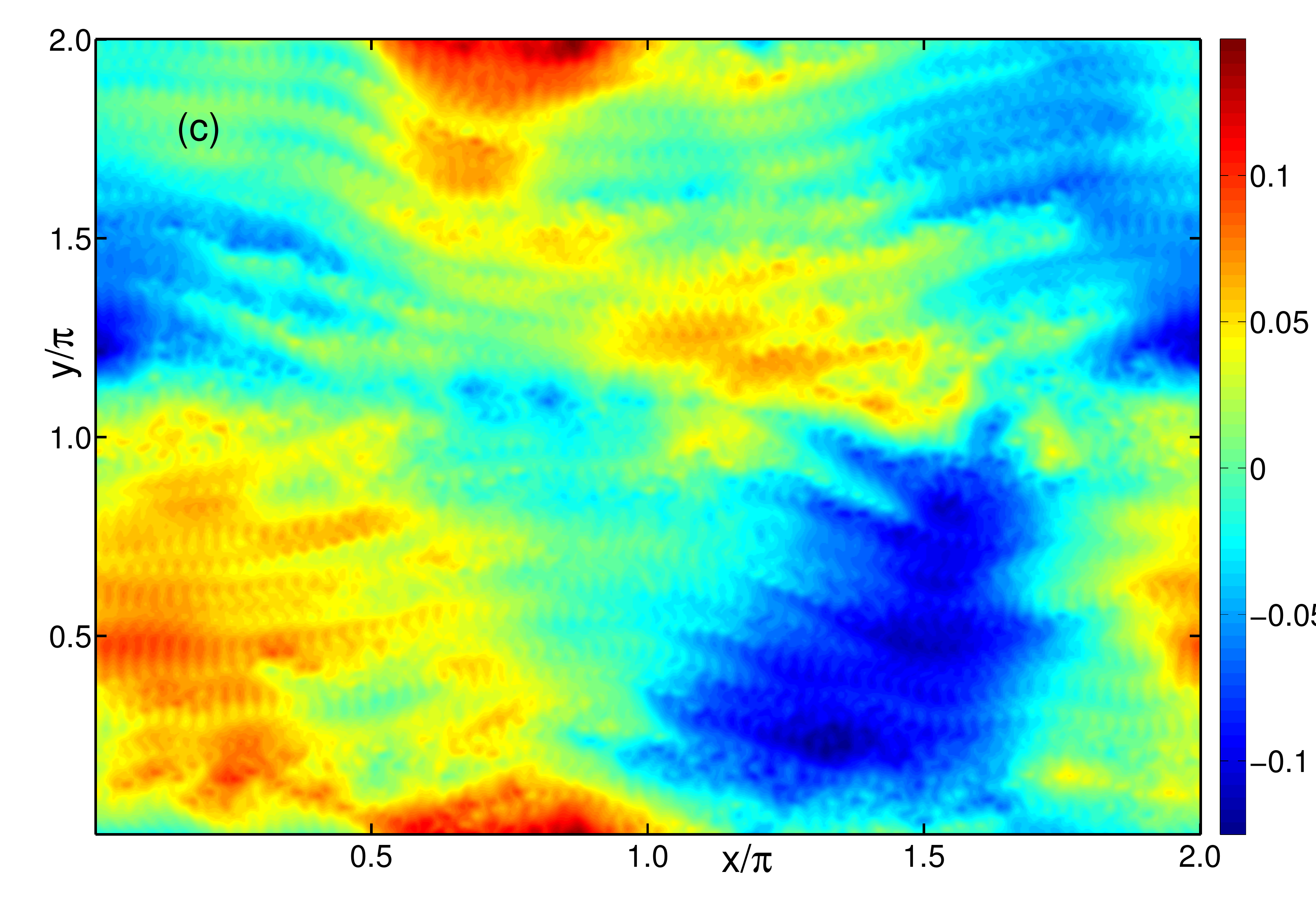}
\includegraphics[width=8cm,height=5cm]{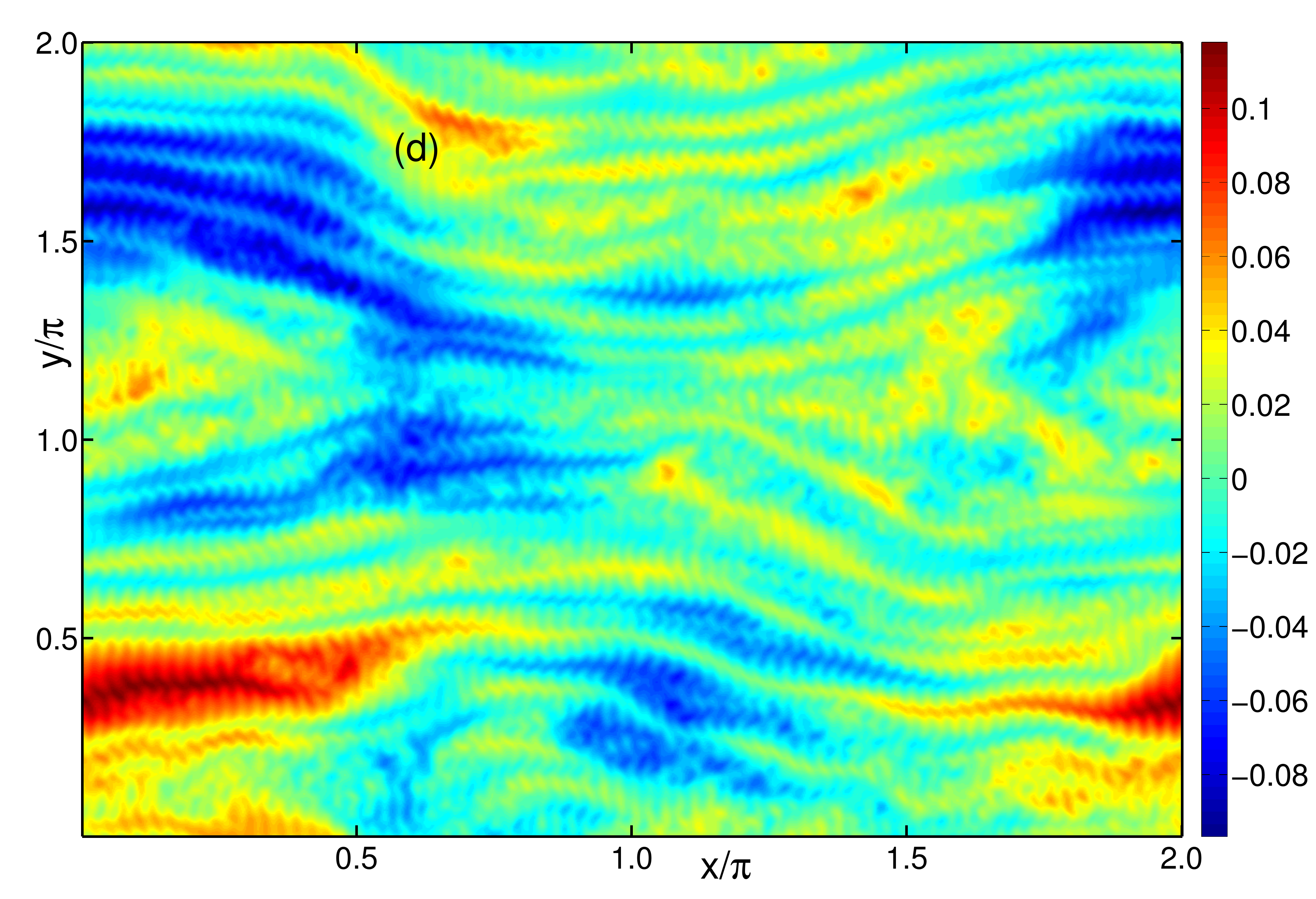}
\caption{(Color online) Pseudocolor plots (a) $\psi$ (run R1), (b) 
$\phi$ (run R1), (c) $\psi$ (run R2), and (d) $\phi$ (run R2).}
\label{fig:snapshot}
\end{figure*}

The Okubo-Weiss parameter $\Lambda$ has been used in several
studies~\cite{okubo,weiss,perlekarnjp} of 2D fluid turbulence to
distinguish between vortical and elongational regions in the
flow. We are not aware of any study of 2D MHD turbulence that
uses $\Lambda$ and $\Lambda_b$ to differentiate, explicitly,
between various flow topologies. (For studies of similar issues
in 3D MHD turbulence, see, e.g., Refs.~\cite{sahoo11,dallas}.) We present
pseudocolor plots of $\Lambda$ (left panels of
Figs.~\ref{fig:overlaid1} and ~\ref{fig:overlaid2}, for runs R1
and R2, respectively) on which we have overlaid the contours of
the stream function $\phi$; in these plots we have zoomed into a
representative, square region in the simulation domain.  (Such
plots for 2D fluid turbulence have been given in
Ref.~\cite{perlekarnjp}.) We give similar pseudocolor plots of
$\Lambda_b$ (right panels of Figs.~\ref{fig:overlaid1} and
~\ref{fig:overlaid2}, for runs R1 and R2, respectively) on which
we have overlaid contours of the magnetic potential $\psi$. From
the left panels of Figs.~\ref{fig:overlaid1} and
~\ref{fig:overlaid2}, we see that vortical regions ($\Lambda >
0$) are associated with centers in the contours of $\phi$,
whereas large-strain-rate regions ($\Lambda < 0$) correspond to
the regions where the contours of $\phi$ have a large curvature.
Similarly, the right panels of Figs.~\ref{fig:overlaid1} and
~\ref{fig:overlaid2} show us that current-dominated regions
($\Lambda_b > 0$) are associated with centers in the contours of
$\psi$, whereas large-magnetic-strain-rate regions  ($\Lambda_b <
0$) correspond to the regions where the contours of $\psi$ have a
large curvature.  We give pseudocolor plots of $\phi$ and
$\psi$ in Figs.~\ref{fig:snapshot} (a) and (b) (for run R1) and
Figs.~\ref{fig:snapshot} (c) and (d) (for run R2), for the
complete simulation domain. 

\subsection{Two-point statistics}
\label{section_twopoint}

\begin{figure*}[htbp]
\includegraphics[width=5.8cm,height=5cm]{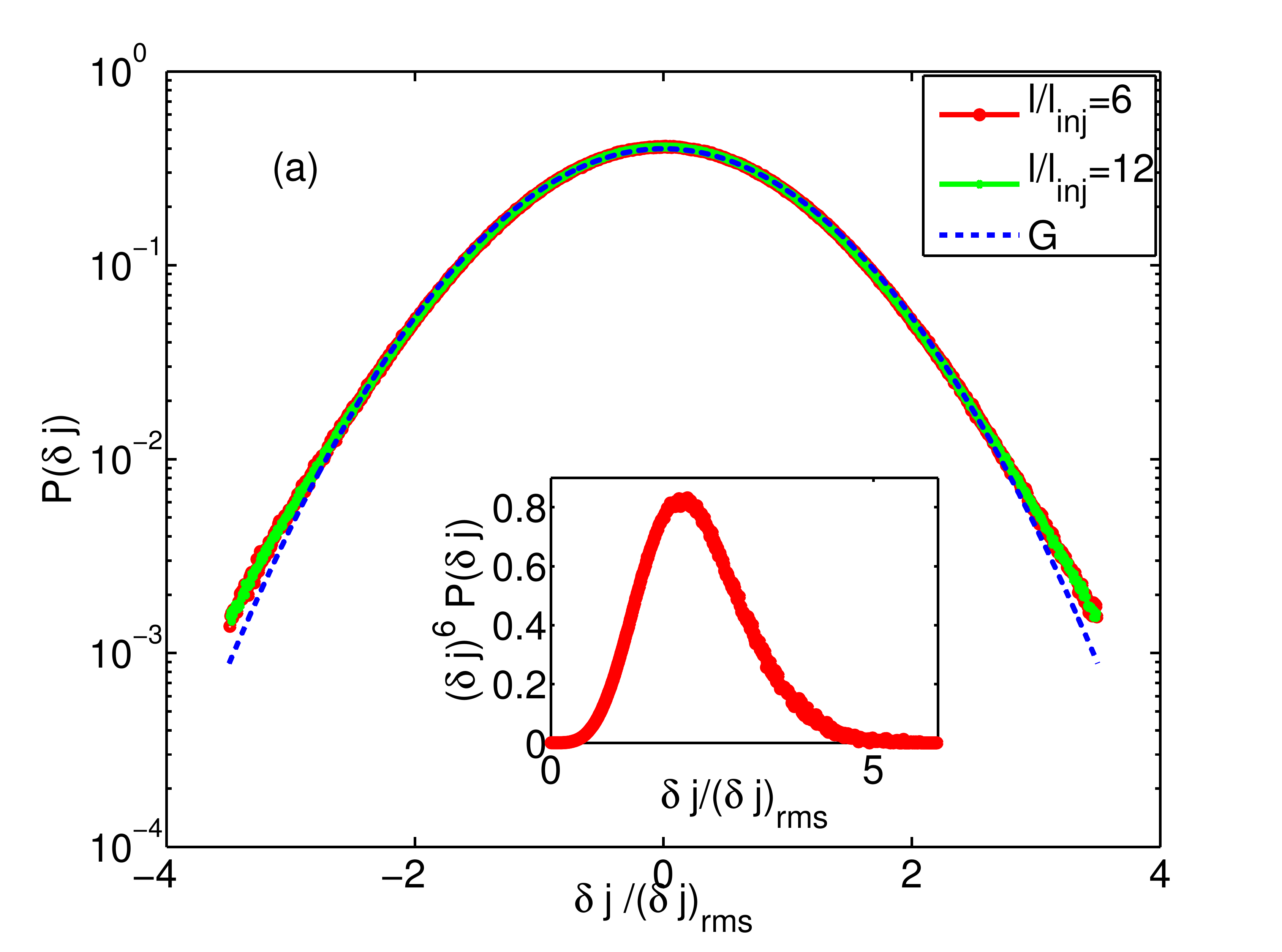} 
\includegraphics[width=5.8cm,height=5cm]{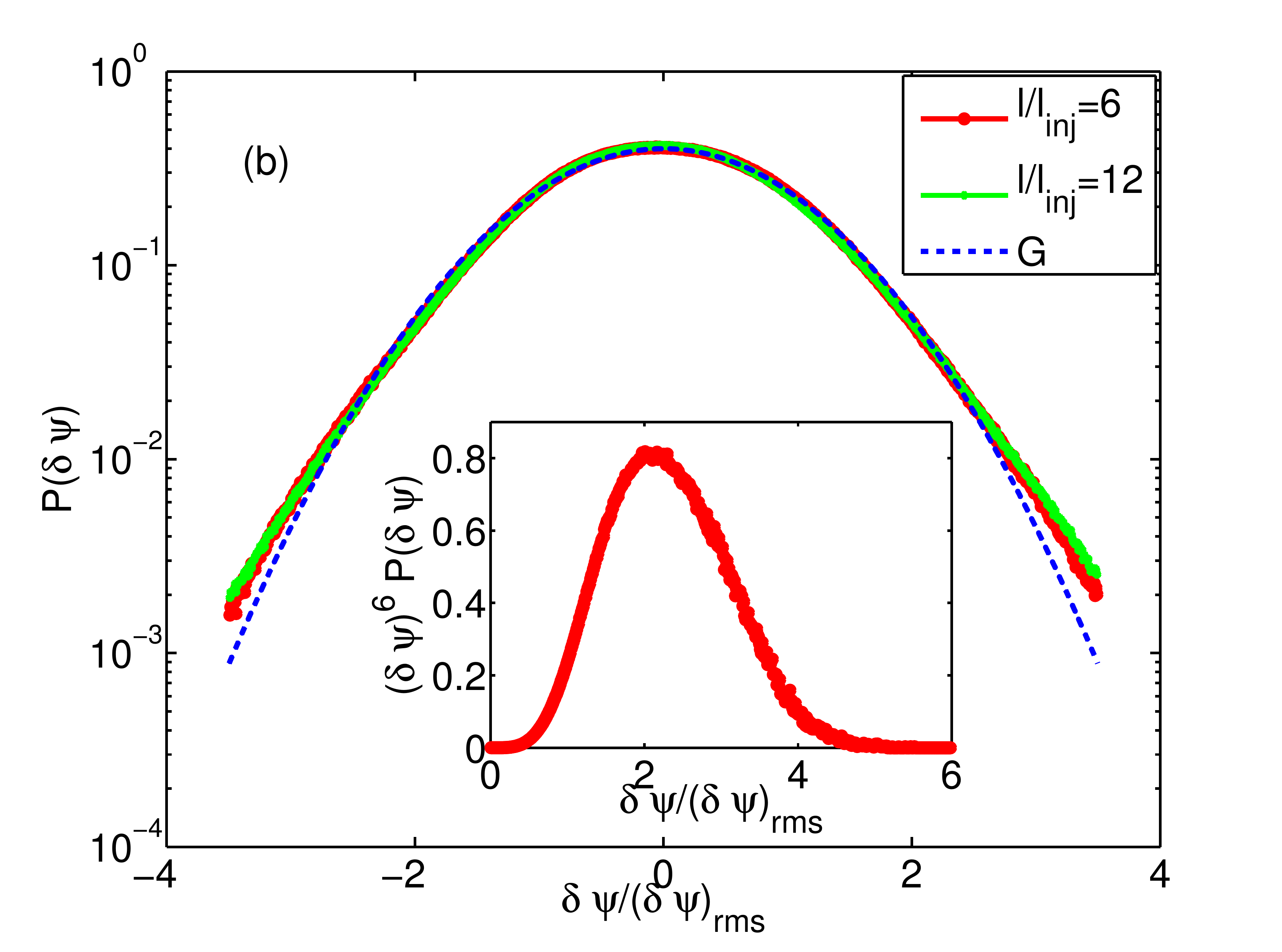}
\includegraphics[width=5.8cm,height=5cm]{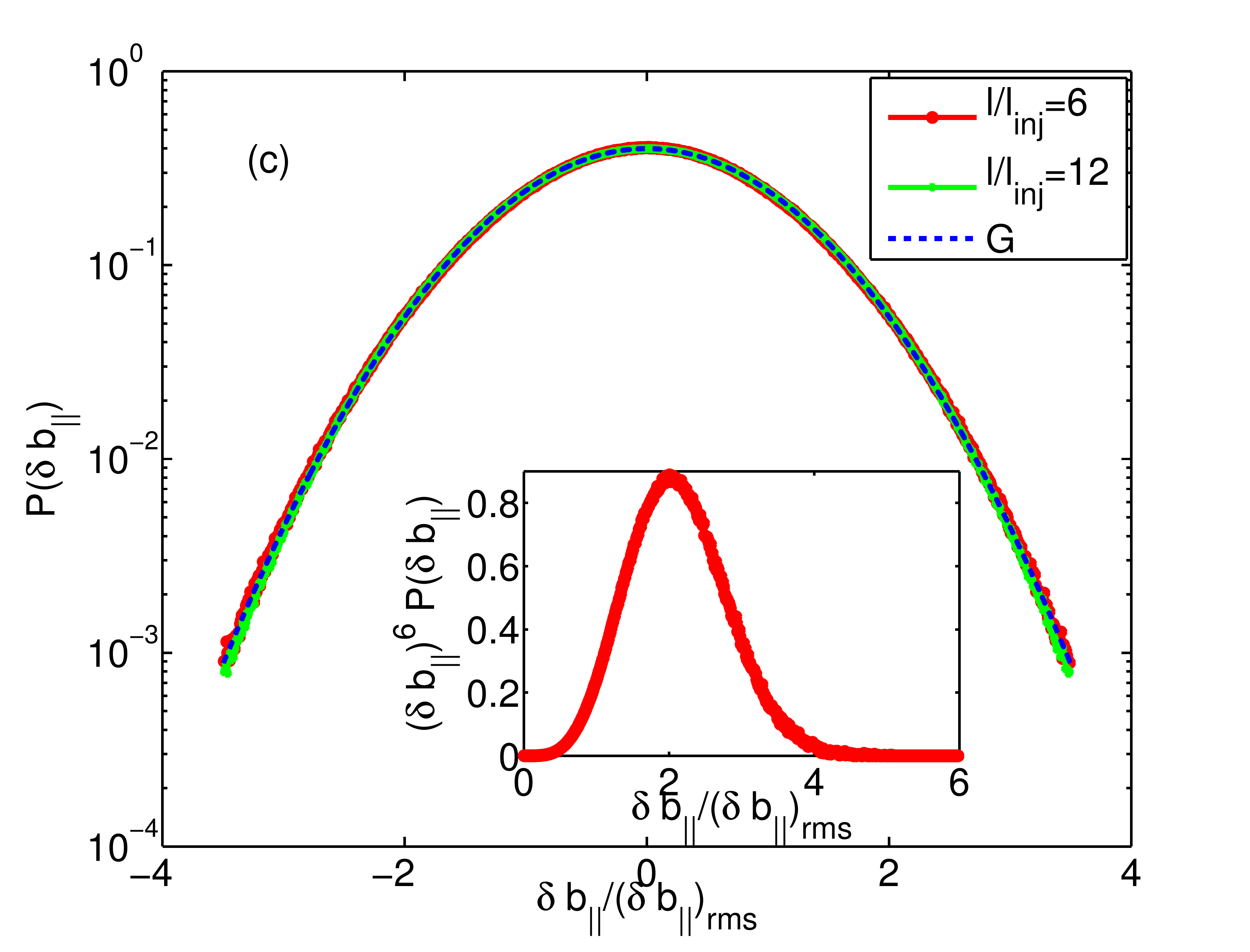} \\
\includegraphics[width=5.8cm,height=5cm]{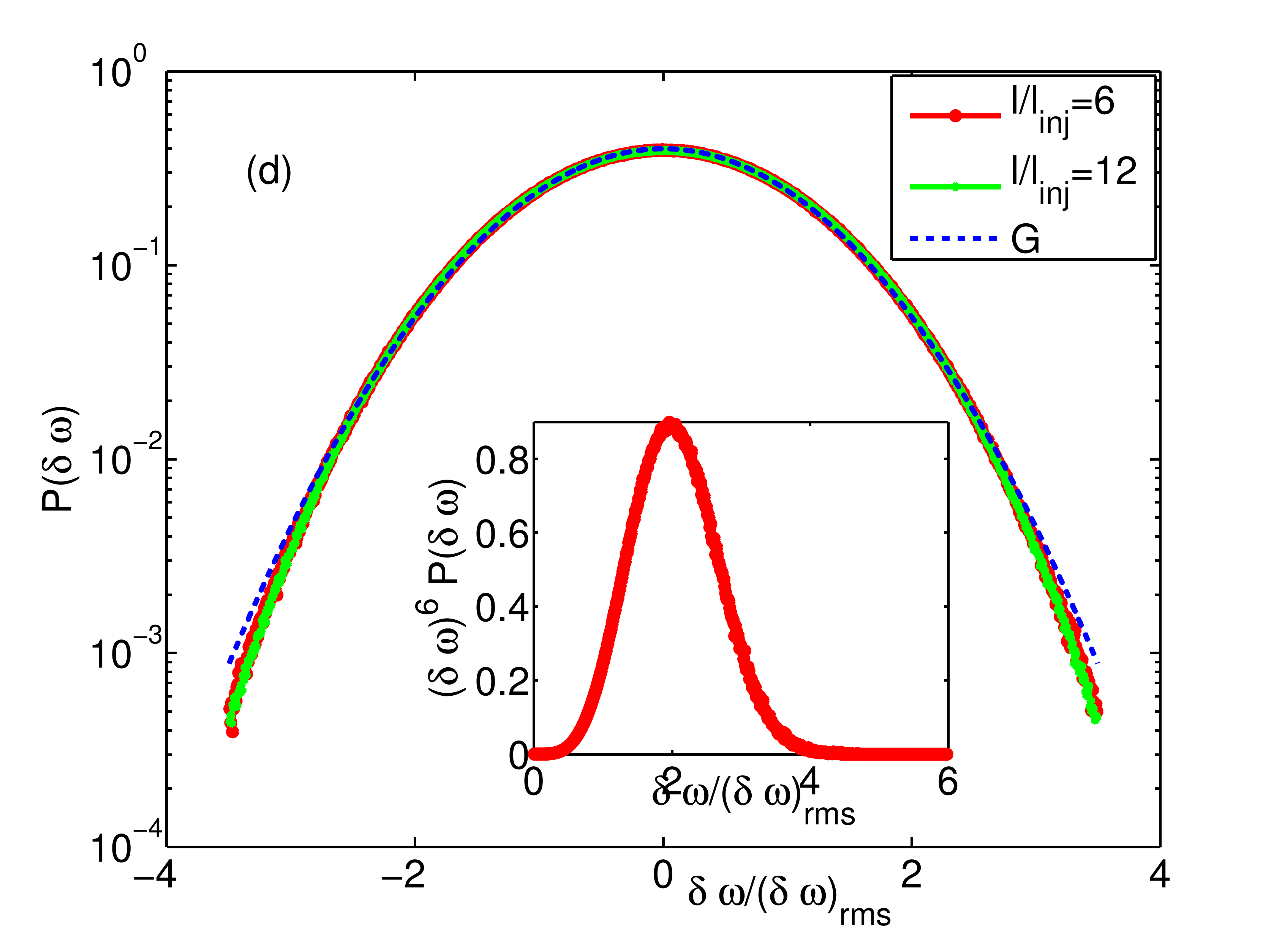}
\includegraphics[width=5.8cm,height=5cm]{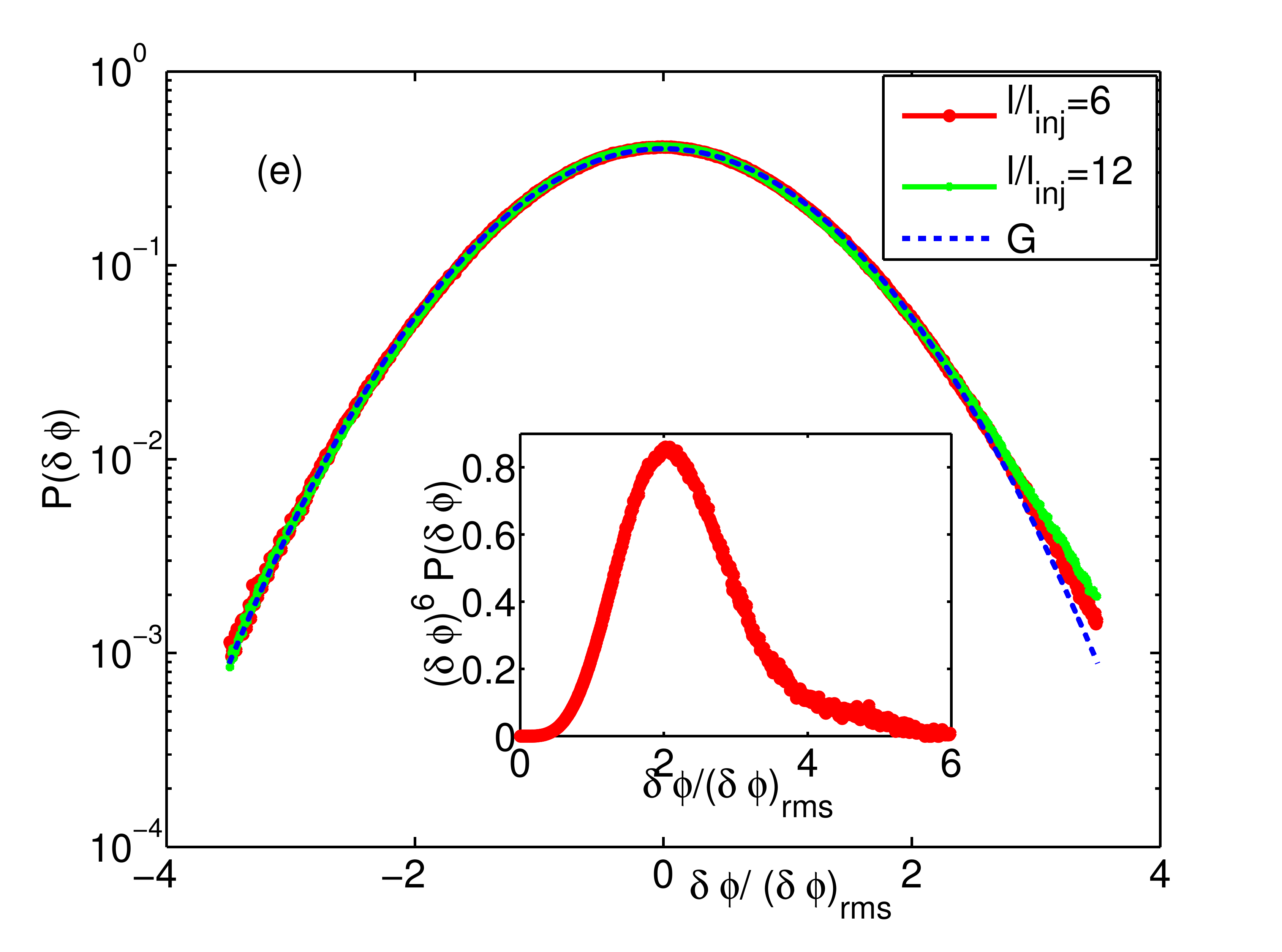}
\includegraphics[width=5.8cm,height=5cm]{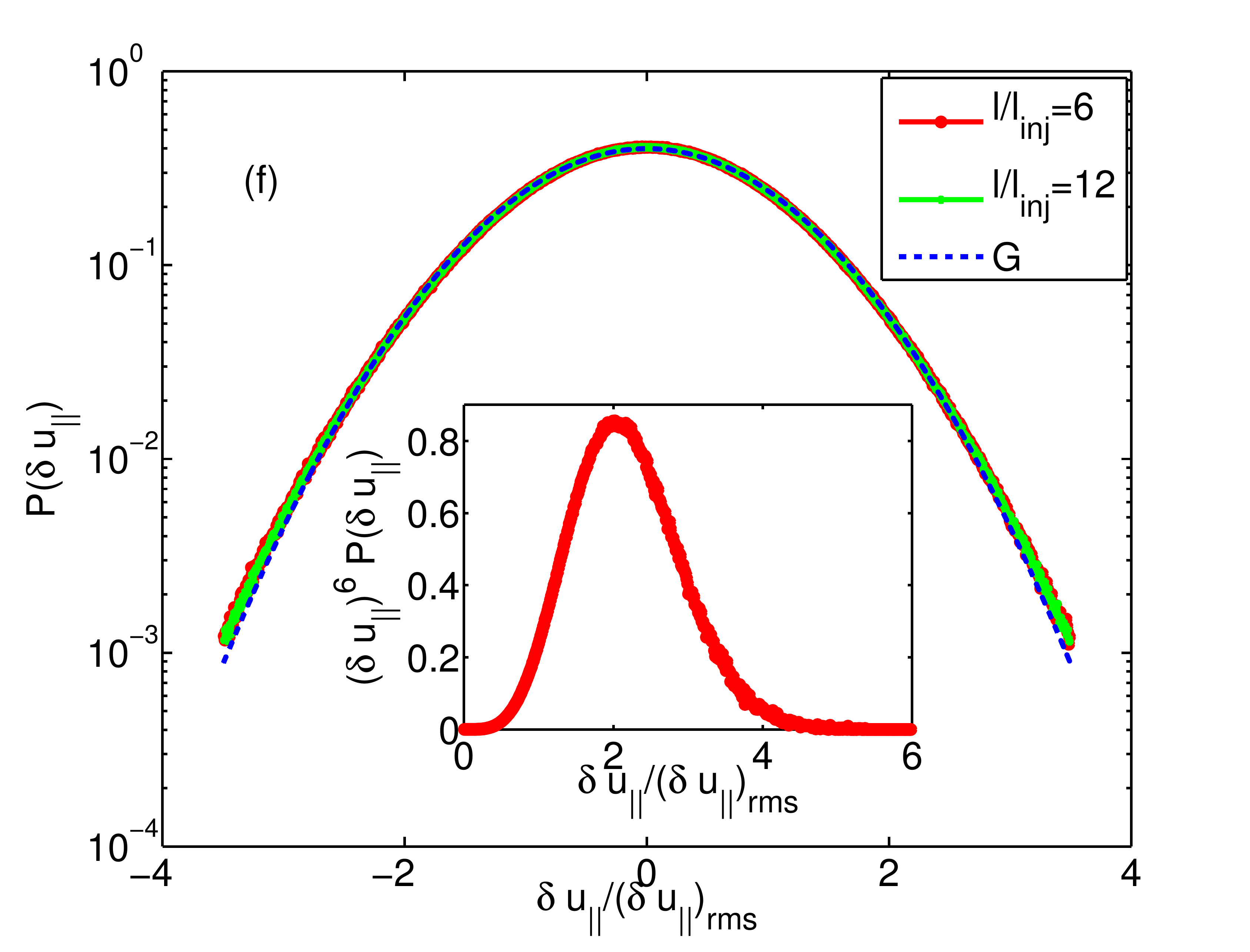}
\caption{(Color online) Plots of PDFs for run R1 of the increments of (a) $\delta j$, 
(b) $\delta \psi$, (c) $\delta b_{||}$, (d) $\delta \omega$, (e) $\delta \phi$, 
and, (f) $\delta u_{||}$, with $l/l_{\rm inj} = 6$ (red curves) and 
$l/l_{\rm inj} = 12$ (green curve). The blue, dashed curves indicate 
Gaussian distributions.} 
\label{fig:pdfs}
\end{figure*}

\begin{figure*}[htbp]
\includegraphics[width=8cm,height=5cm]{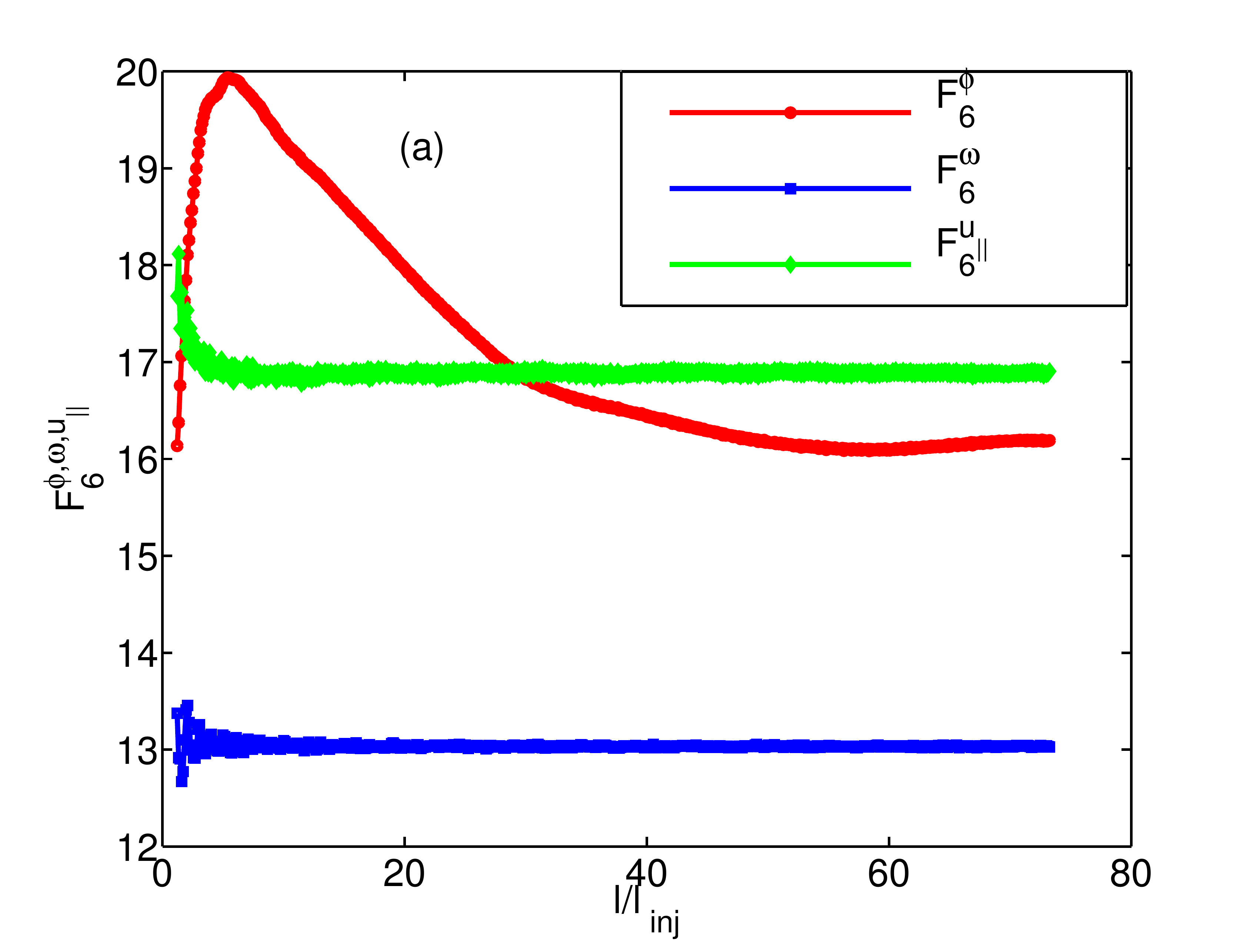}
\includegraphics[width=8cm,height=5cm]{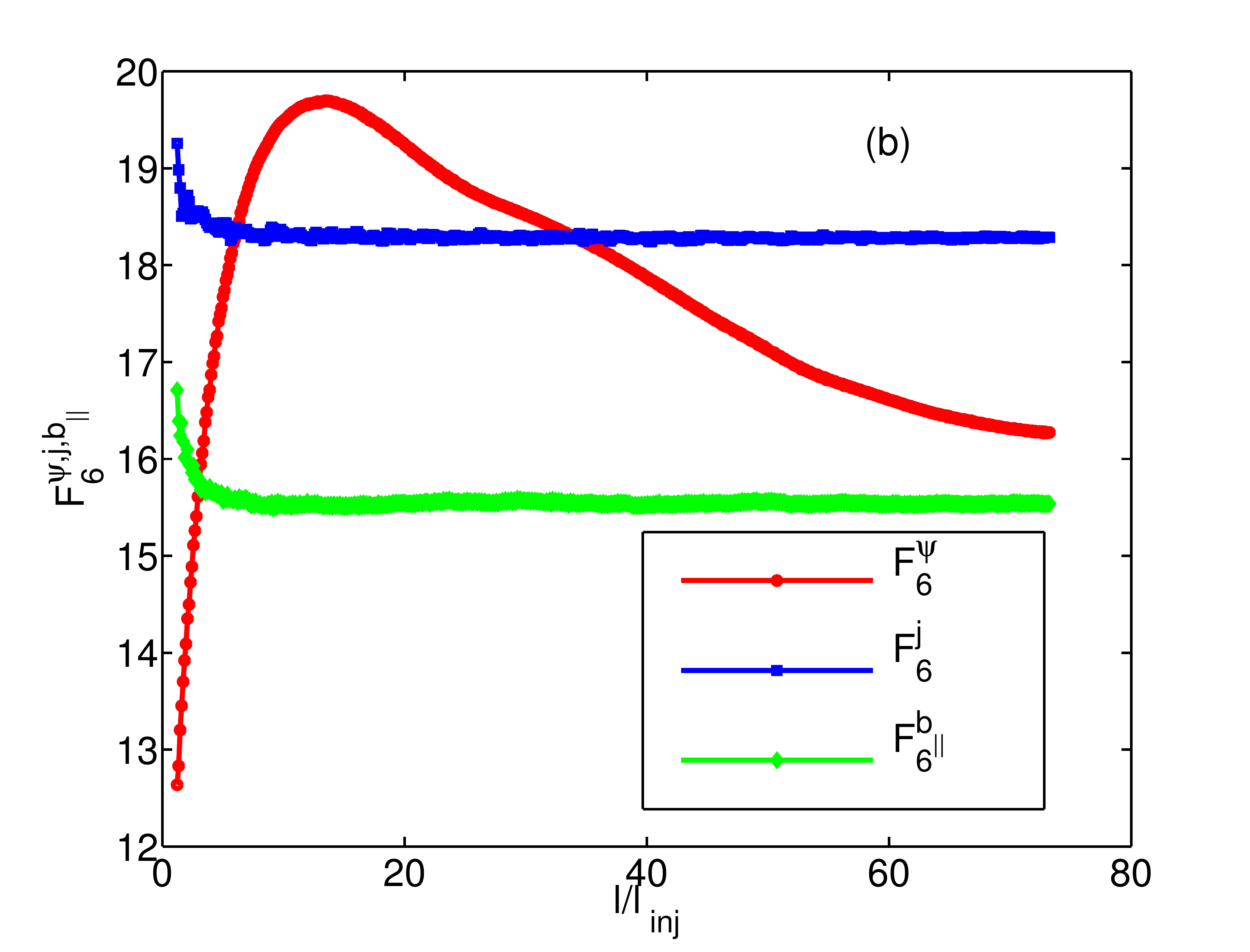} \\
\includegraphics[width=8cm,height=5cm]{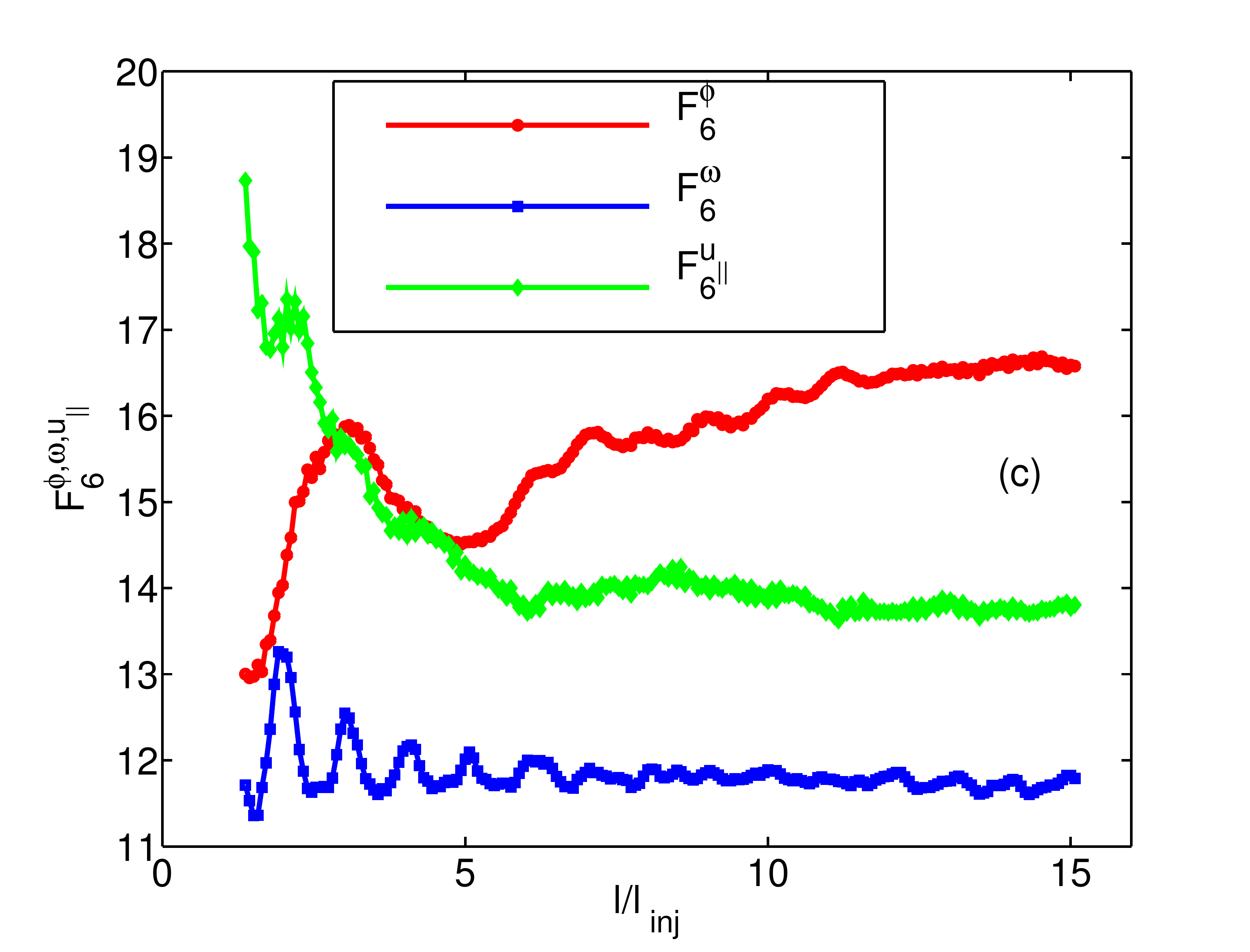}
\includegraphics[width=8cm,height=5cm]{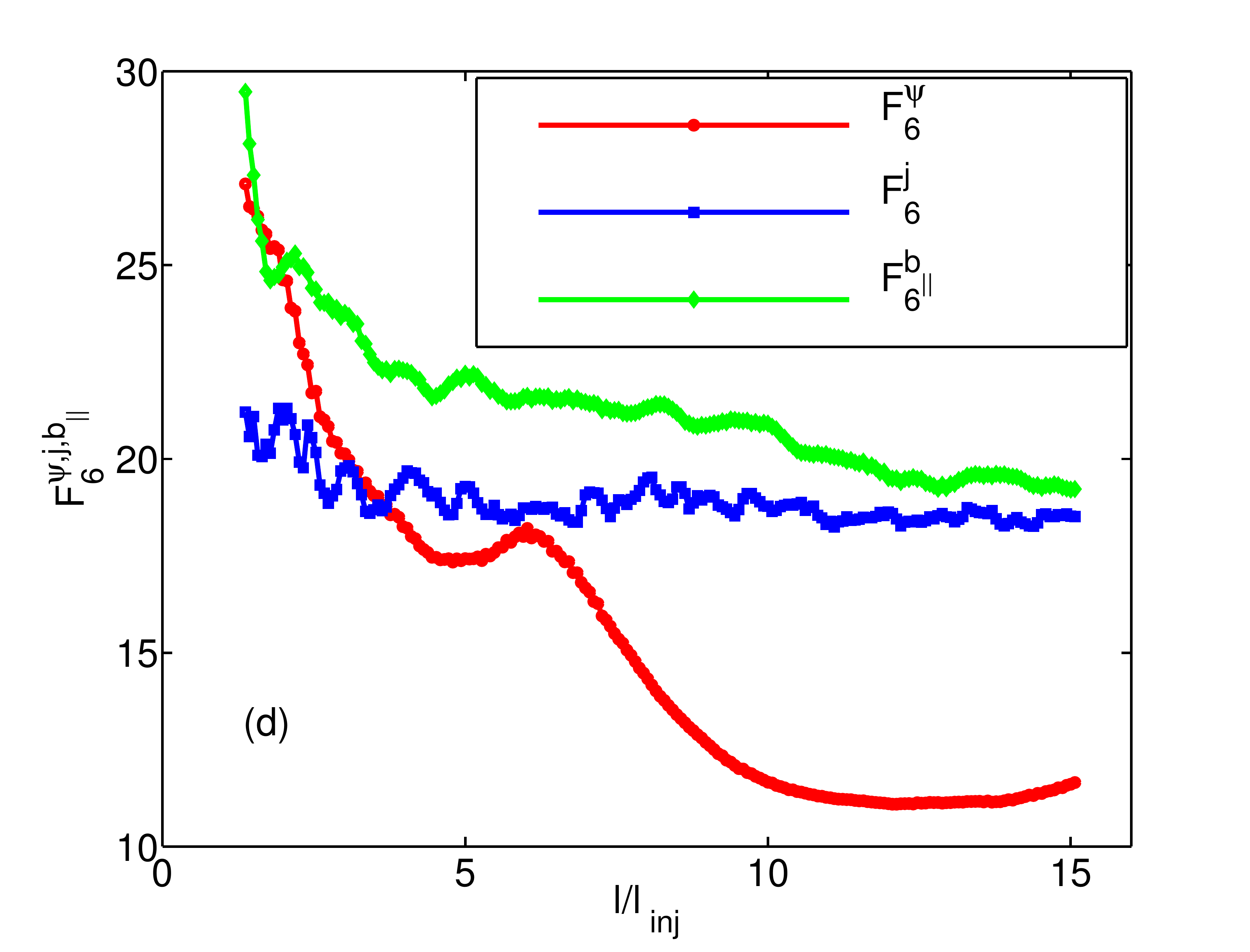}
\caption{(Color online) Plots of the hyperflatnesses 
versus the scaled length $l/l_{\rm inj}$ for the
following increments: (a) $\delta \phi$ (red curve),
$\delta \omega$ (blue curve), $\delta u_{||}$ (green curve)
(for run R1), (b) $\delta \psi$ (red curve), $\delta j$
(blue curve), $\delta b_{||}$ (green curve) (for run R1),
(c) $\delta \phi$ (red curve), $\delta \omega$ (blue
curve), $\delta u_{||}$ (green curve) (for run R2), and (d)
$\delta \psi$ (red curve), $\delta j$ (blue curve), $\delta b_{||}$
(green curve) (for run R2).}
\label{fig:hyper}
\end{figure*}

In Figs.~\ref{fig:pdfs} (a)-(f), we plot PDFs of the field
increments (a) $\delta j$, (b) $\delta \psi$, (c) $\delta
b_{||}$, (d) $\delta \omega$, (e) $\delta \phi$, and (f)
$\delta u_{||}$, for $l/l_{\rm inj} = 6$ (red curves) and
$l/l_{inj} = 12$ (green curves).  These PDFs lie very close
to Gaussian ones, which are shown by dashed blue curves;
small, non-Gaussian deviations appear in the PDF tails. To
quantify the scale dependence of these deviations, we
obtain the hyperflatness $F_6^a(l) \equiv
S_6^a(l)/(S_2^a(l))^3$ for all fields $a$. In the insets of
Figs.~\ref{fig:pdfs} (a)-(f), we show plots of $\delta j^6
{\mathcal P}(\delta j)$ versus  $\delta j$, etc., to
demonstrate that, at least up to order $6$, our structure
functions are reliable.  We present plots versus
$l/l_{\rm inj}$ of the hyperflatnesses of the field increments
in the left and right panels of Fig.~\ref{fig:hyper} for (a)
$\delta \phi$ (red curve), $\delta \omega$ (blue curve),
and $\delta u_{||}$ (green curve) and (b) $\delta \psi$ (red
curve), $\delta j$ (blue curve), and $\delta b_{||}$ (green
curve). From these plots we see that, in run R1,
$F_6^{\phi}(l)$ and $F_6^{\psi}(l)$ (red curves in
Figs.~\ref{fig:hyper} for (a) and (b), respectively) lie
slightly above the Gaussian value $15$ for $l/l_{\rm inj} >
35$, they increase gently for lower values of $l/l_{\rm
inj}$, indicating an enhancement of small-scale
intermittency, and decrease again towards the Gaussian
value after going through at maximum at which these
hyperflatnesses are $\simeq 20$.  In contrast,
$F_6^{u_{||}}(l)$ and $F_6^{b_{||}}(l)$ (green curves in
Figs.~\ref{fig:hyper} for (a) and (b), respectively) do not
show significant scale dependence; the former lies $\simeq
13 \%$ above the Gaussian value, whereas the latter is only
a few percent above this. $F_6^{\omega}(l)$ and
$F_6^{j}(l)$ (blue curves in Figs.~\ref{fig:hyper} for (a)
and (b), respectively) also do not show significant scale
dependence; the former lies $\simeq 13 \%$ below the
Gaussian value, whereas the latter is $\simeq 20 \%$ above
this. Given that the spectra for runs R1 and R2 are
different, it is not surprising that the scale dependences
of the hyperflatnesses $F_6^a(l)$ are different for these
runs too, as we can see by comparing the plots in
Figs.~\ref{fig:hyper} (a) and (b), for run R1, with their
counterparts in Figs.~\ref{fig:hyper} (c) and (d), for run
R2.

\begin{figure*}
\includegraphics[width=5.8cm,height=5cm]{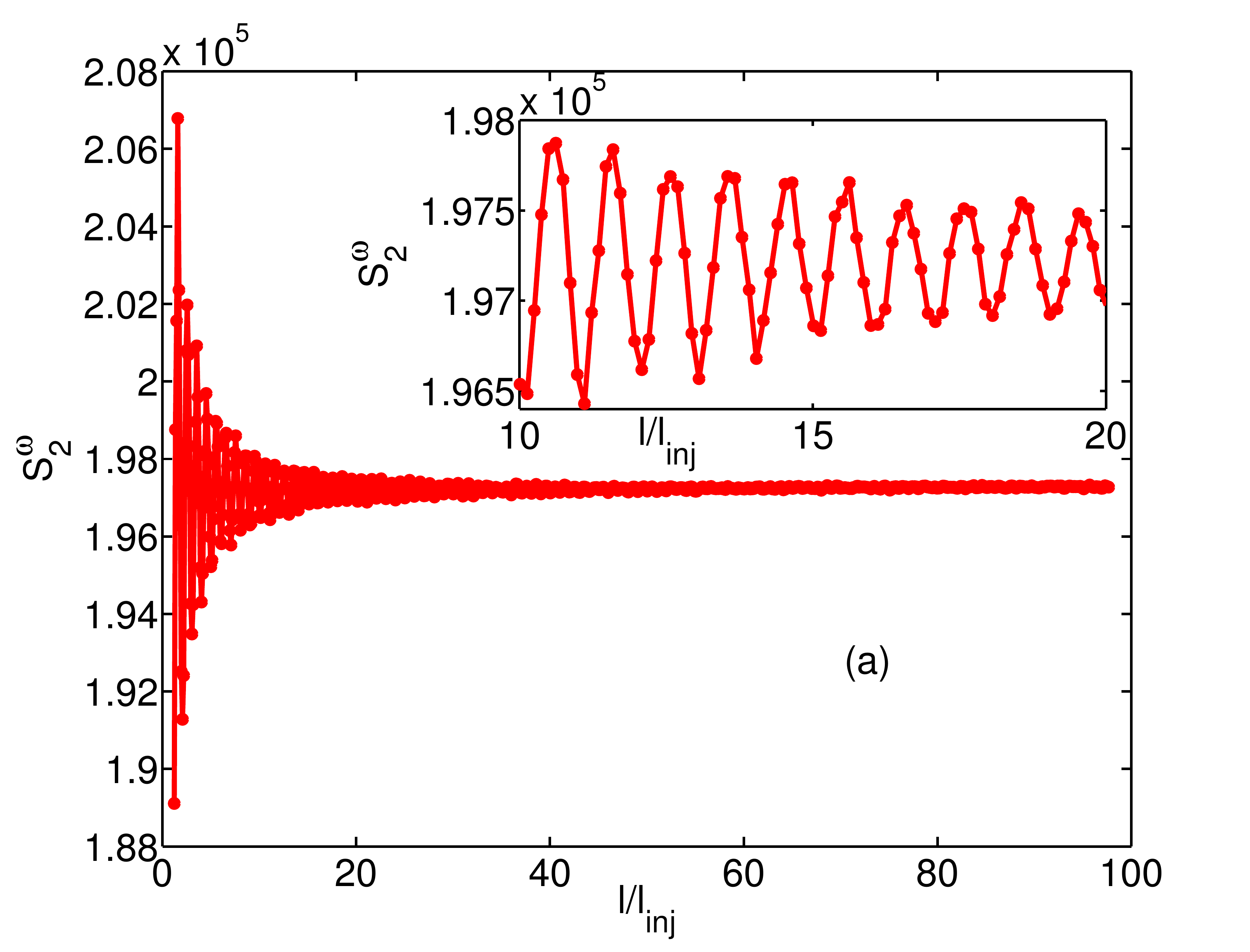}
\includegraphics[width=5.8cm,height=5cm]{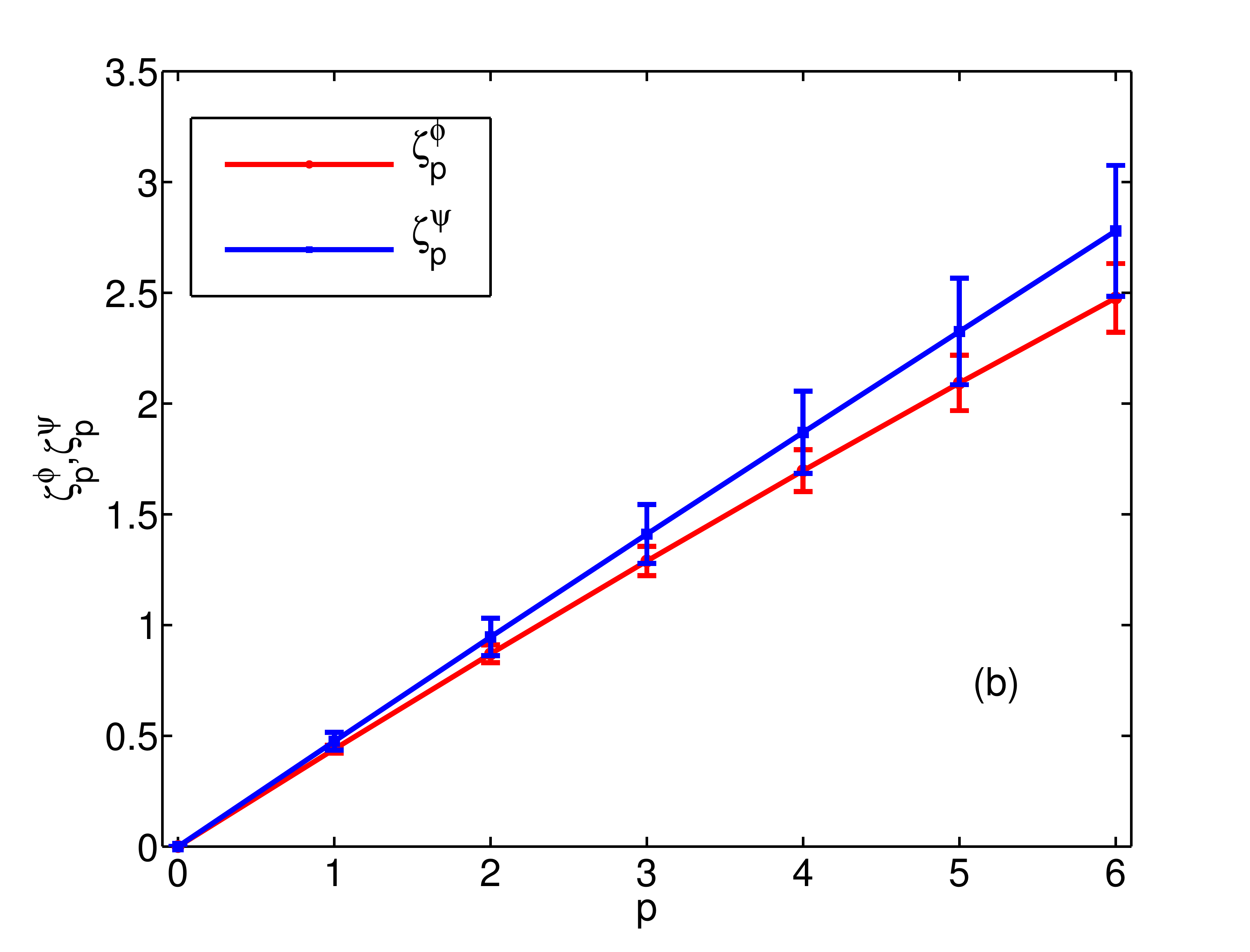}
\includegraphics[width=5.8cm,height=5cm]{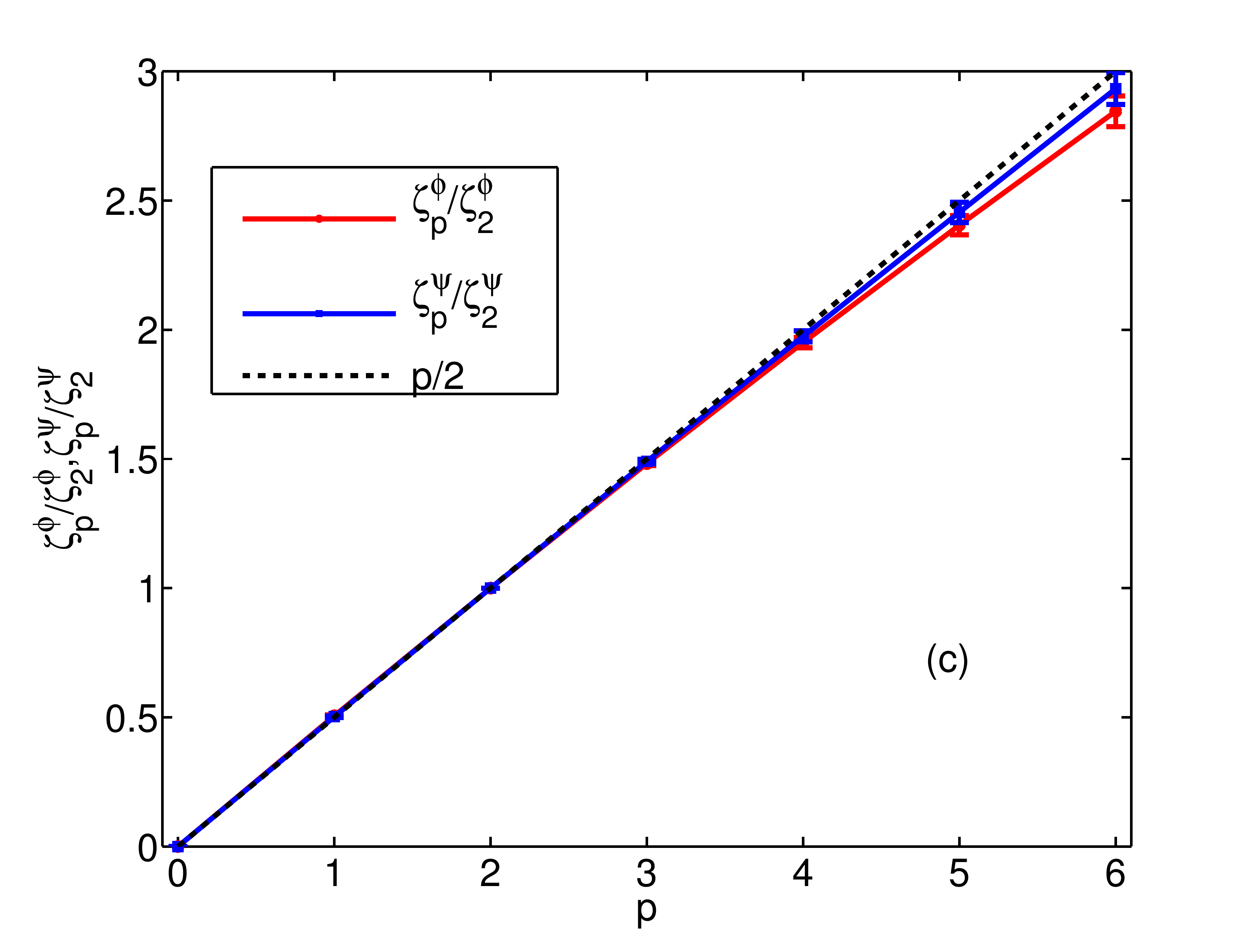}
\caption{(Color online) Plots, for run R1, of 
(a) $S_2^{\omega}$ versus $l/l_{\rm inj}$ (in the inset
we zoom into the region $10l_{\rm inj} \le l \le 20l_{\rm inj}$), 
(b) multiscaling exponents versus $p$ for $\phi$ (red curve) and 
$\psi$ (blue curve), without extended self similarity (ESS - see
text), and (c) multiscaling exponents ratios versus $p$ for $\phi$ 
(red curve) and $\psi$ (blue curve), with ESS. For these multiscaling 
exponent we use data for the structure functions in the range $10 \le l/l_{\rm inj} \le 30$.}
\label{fig:strfn}
\end{figure*}

Note that there are well-defined, damped oscillations in
$F_6^{\omega}(l)$ and $F_6^{j}(l)$, at small values of
$l/l_{\rm inj}$ (these show up especially clearly on the
scales of  Figs.~\ref{fig:hyper} (c) and (d)). The distance
between successive maxima in these oscillations is $\simeq
l_{\rm inj} = 2\pi/k_{\rm inj}$. Such damped oscillations
also appear in plots of structure functions; we show an
illustrative plot for $S_2^{\omega}(l)$ in
Fig.~\ref{fig:strfn} (a). The origin and form of these
oscillations can be understood easily for second-order
structure function of field $a$ because it is related, via Fourier
transformation, to the spectrum $|a(k)|^2$. We show
this explicitly below for $S_2^{\omega}(l)$. 

We consider homogeneous, isotropic, turbulence, so $S_2^{\omega}$
depends only on $l = |{\bf l}|$, therefore,
\begin{eqnarray}
S_2^{\omega}(l) & \sim &  \int_0^{k_{UV}} |\omega(k)|^2 [1 -
J_0(kl)] dk,  \nonumber \\
S_2^{\omega}(l) & \sim &  \int_0^{k_{UV}} k^2 [1 - 
J_0(kl)] dk, \nonumber \\
S_2^{\omega}(l) & \sim & \frac{1}{l^3} \int_0^{k_{UV} l} x^2
[1-J_0(x)]dx,
\end{eqnarray}
where we control the ultra-violet (UV) divergence of the
integrals by using a UV cutoff $k_{UV}$ and make the assumption
$|\omega(k)|^2 \sim k^2$, which is consistent with the spectrum in
Fig.~\ref{fig:spectra} for run R1. (For run R2, $|\omega(k)|^2 \sim
k^{2.3}$, so the subsequent steps cannot be carried out
analytically; however, the oscillations are more pronounced
than they are in the case $|\omega(k)|^2 \sim k^2$.) We obtain,
finally,
\begin{widetext}
\begin{equation}
S_2^{\omega}(l)  \sim  \frac{1}{l^3} \lbrack
\frac{1}{6}k_{UV}l(2k_{UV}l(k_{UV}l-3J_1(k_{UV}l)) + 3
\pi J_1(k_{UV}l)H_0(k_{UV}l) - 3\pi
J_0(k_{UV}l)H_1(k_{UV}l)) \rbrack,
\label{eq:stf1}
\end{equation}
\end{widetext}
where $J_n$ and $H_n$, with $n = 0$ or $1$, denote, respectively,
Bessel and Struve functions~\cite{dlmf}, which oscillate in 
a manner that is consistent with the plot in Fig.~\ref{fig:strfn}. (a)
Similar arguments for the second-order, logitudinal velocity structure 
function (with $ |u(k)|^2 \sim k^0$ in Fig.~\ref{fig:spectra} for run R1)
yield

\begin{widetext}
\begin{equation}
S_2^{u_{||}}(l)  \sim  \frac{1}{l} \lbrack k_{UV}l -\frac{1}{2}(\pi J_1(k_{UV}l)
H_0(k_{UV}l) +  J_0(k_{UV}l)(2-H_1(k_{UV}l))) \rbrack. \\
\label{eq:stf2}
\end{equation}
\end{widetext}

The structure functions for $\psi$ and $\phi$ do not show such oscillations
because the Fourier integrals are UV convergent.  

The oscillations appear much more clearly in $S_2^{\omega}$
than in $S_2^{u_{||}}$ because, for large $k_{UV} l \gg 1$,
$S_2^{\omega} \sim k_{UV}^3 - k_{UV}^2 \frac{3J_1(k_{UV}l)}{l}$,
whereas $S_2^{u_{||}} \sim k_{UV} - \frac{1}{2l}(\pi J_1(k_{UV}l)
H_0(k_{UV}l) + J_0(k_{UV}l)(2-H_1(k_{UV}l)))$. The relative
strengths of these oscillations is governed by the ratio of the
coefficients of the second terms in  $S_2^{\omega}$ and
$S_2^{u_{||}}$ is proportional to $k_{UV}^2/l$, which is large
because $k_{UV}$ can be taken to be $k_{\rm inj}$ (this is
consistent with the period of oscillations that we see in the
inset of Fig.~\ref{fig:strfn} (a)). Thus, we expect the oscillations in
$S_2^{\omega}$ to be more conspicuous than their counterparts in
$S_2^{u_{||}}$. From the spectra presented in
subsection~\ref{section_spectra}, we know that the constants of 
proportionality, in Eqs.~(\ref{eq:stf1}) and (\ref{eq:stf2}) for $S_2^{\omega}$ and $S_2^{u_{||}}$
are of the same order of magnitude.

Given the oscillations in $S_2^{\omega}$ and $S_2^{u_{||}}$, and the lack of
significant scale dependence in the plots of $F_6^{\omega}$ and $F_6^{u_{||}}$
(Figs.~\ref{fig:hyper} (a)-(d)), we do not expect noticeable multiscaling in
vorticity and velocity structure functions in the inverse-cascade range.
However, given the mild scale dependence in the plots of $F_6^{\phi}$ and
$F_6^{\psi}$, for run R1 (red curves in Figs.~\ref{fig:hyper} (a)-(d)), we have
explored power-law fits of the form  $S_p^{\phi}(l) \sim l^{\zeta^\phi_p}$ and
$S_p^{\psi}(l) \sim l^{\zeta^\psi_p}$, in the range $ 10 \leq l/l_{\rm inj}
\leq 30$, and have obtained from there the multiscaling exponents $\zeta^\phi_p$ and
$\zeta^\psi_p$, which we plot as red and blue curves, respectively, in
Fig.~\ref{fig:strfn}(b). For studies of multiscaling in 3D MHD see Refs.~\cite{sahoo11,3dmhd,mhdshell}. 
The error bars in Fig.~\ref{fig:strfn}(b) improve
slightly if we use the extended self-similarity (ESS) procedure to extract the
multiscaling exponent ratios $\zeta^\phi_p/\zeta^\phi_2$ and
$\zeta^\psi_p/\zeta^\psi_2$, as we show in  Fig.~\ref{fig:strfn}(c). [In the  (ESS)
procedure~\cite{ess} we extract multiscaling exponent ratios from power-law
ranges in plots of $S_p^{\phi}(l)$ versus $S_2^{\phi}(l)$ and $S_p^{\psi}(l)$
versus $S_2^{\psi}(l)$.] We obtain error bars by carrying out local-slope
analyses in the power-law ranges of these structure functions. The deviations of the 
multiscaling exponent ratios from a simple-scaling straight line (see the dashed line in Fig.13c) are
very small. Thus, only very-high-resolution DNS can settle whether such multiscaling
exists at all; such DNSs lie beyond the scope of this study.

\section{conclusions}

We have presented the most comprehensive study of the statistical
properties of homogeneous, isotropic 2D MHD turbulence in the
inverse-cascade regime. Our work has used very long simulations
to make sure that we obtain statistically steady states.
Furthermore, we have calculated many more statistical properties
compared to earlier studies of 2D MHD
turbulence~\cite{biskamp,celani}. We have shown that these statistical
properties are different for runs R1 and R2, i.e., these
properties depend on the friction, viscosity, and magnetic-diffusivity 
coefficients that distinguish these runs.

The spectra for various fields are different in runs R1
(Fig.~\ref{fig:spectra}) and R2  (Fig.~\ref{fig:spectra2}), as we have discussed in
detail above. In particular, the exponents, which characterize
the power-law behaviors of these spectra in the inverse-cascade
range, are different in runs R1 and R2; and these exponents are
different from the dimensional prediction  $E^u(k)\sim E^b(k)\sim
k^{-1/3}$. The study of Ref.~\cite{biskamp} yields energy spectra
that are consistent with $E^u(k)\sim k^{1/3}$ and $E^b(k)\sim
k^{-1/3}$; by contrast, the results of Ref.~\cite{celani} imply
$E^u(k)\sim k^{1/3}$ and $E^b(k)\sim k^{0}$ 
(Eqs.~\ref{eq:dimensions1},~\ref{eq:dimensions2},~\ref{eq:dimensions3},~\ref{eq:dimensions4}). 
Our run R1 yields energy spectra that are consistent with $E^u(k)\sim k^{0}$ and
$E^b(k)\sim k^{-1/3}$; their analogs for run R2 are $E^u(k)\sim
k^{1/3}$ and $E^b(k)\sim k^{-1/3}$. 

The spectral exponents of Ref.~\cite{biskamp} agree with 
those of our run R2. However, a careful comparison of our 
spectra (Fig.~\ref{fig:spectra2}) with theirs (Fig.~\ref{fig:spectra2} in 
Ref.~\cite{biskamp}) reveals important differences at small 
values of $k$: our energy spectra increase slightly at very small 
$k$, because we do not use friction in run R2, whereas those of 
Ref.~\cite{biskamp} fall in this range, in a manner that suggests
a friction term, but this is not mentioned explicitly in their paper. Note
that Ref.~\cite{biskamp} employs hyperviscosity (fourth power
of the Laplacian), whereas we use conventional viscosity in 
run R2. Therefore, we might expect, {\it a priori}, that the
spectra of Ref.~\cite{biskamp} may differ from those of run R2
only at large values of $k$ (from the bottleneck region around
$k_{\rm inj}$ and beyond).

The spectral exponents of Ref.~\cite{celani} agree with 
those of our run R2 for the velocity field but not for 
the magnetic field (or its potential). The low-$k$ form of 
their spectra suggests that they employ a friction term or
some other mechanism for large-scale dissipation. They also 
report a power-law form, $E^u(k)\sim k^{-5/3}$, in the
forward-cascade regime, which we do not study here
(we choose a large value of $k_{\rm inj}$ so that we have a large
inverse-cascade range).

Thus, spectral features seem to depend in detail on the precise
dissipation or friction terms that are used in a DNS and perhaps
also on the details of the forcing (e.g., whether both or one of
the velocity or magnetic fields are forced). A full elucidation
of such dependences must await very-high-resolution and long
DNSs. Such high-resolution DNSs might also be required to 
remove the mild ripples in the pseudocolor plots of Figs.~\ref{fig:snapshot} (c) and (d)
from run R2, which does not use hyperviscosity, which 
suppresses these ripples (cf. Figs~\ref{fig:snapshot} (a) and (b) for run R1).

We have presented PDFs of $\omega$, $j$, $\phi$, and $\psi$ in
Fig.(\ref{fig:one_point_pdf}) for the run R1; we find that these PDFs are
similar for run R2, so we do not give them here. All these PDFs
are close to Gaussian ones, but they show slight deviations from
Gaussians in their tails. We have quantified such deviations by
calculating the flatnesses of these PDFs. We find, in particular,
that the PDFs of $\omega$ and $j$ deviate from Gaussian
distributions in two different ways: the PDF of $\omega$ falls
more steeply than a Gaussian, the PDF of $j$ falls less steeply
than a Gaussian. The PDFs of $\phi$ and $\psi$ hardly deviate
from Gaussian distributions; this last result is consistent with
that of Ref.~\cite{celani}. 

We have quantified the degree of alignment
between $\omega$ and $j$ by obtaining the PDFs of
$\cos(\beta_{u,b})$ and $\cos(\beta_{\omega,j})$. From these PDFs
we see that (a) $\omega$ and $j$ are either parallel or
anti-parallel with equal probability and (b) the PDF of
$\cos(\beta_{u,b})$ has a minimum at $0$, i.e., there is a low
probability of orthogonality of $\bf u$ and $\bf b$, and is
symmetrical about this minimum, i.e., there is equal probability
of alignment and anti-alignment.  The latter result is similar to
that obtained in Ref.~\cite{alignment} for decaying 2D MHD
turbulence.

Calculations of $\Lambda$ and $\Lambda_b$ have not been attempted
earlier for 2D MHD. However, there have been many studies of
$\Lambda$ for 2D fluid
turbulence~\cite{okubo,weiss,perlekarnjp,persistenceprl,aguptanew}
and one for $\Lambda$ in fluid turbulence with polymer
additives~\cite{anupampaper}. Our PDFs for $\Lambda$ and its
magnetic analog $\Lambda_b$ (Figs.~\ref{fig:lambda1} 
(a) and ~\ref{fig:lambda2} (a)) are
qualitatively similar to their fluid-turbulence counterparts
(see, e.g., Fig.(7) in Ref.~\cite{perlekarnjp}) insofar as they
have cusps at the origin and have tails that can be fit to
exponential forms. Our joint PDFs of $\Lambda$ and $\Lambda_b$,
for both runs R1 and R2, are sharply peaked at the origin and
display ridges that separate the four quadrants in the
$\Lambda-\Lambda_b$ plane. By using pseudocolor plots $\Lambda$
and $\Lambda_b$, with superimposed contour lines of $\phi$ and
$\psi$, respectively, we have shown that $\Lambda > 0$ in
vortical regions, $\Lambda < 0$ in extensional regions,
$\Lambda_b > 0$ in current-dominated regions, and $\Lambda_b < 0$
in magnetic-strain-dominated regions

The field-increment PDFs have been studied earlier for the field
$\psi$ in the Ref.~\cite{celani}, which has reported a small, but
finite, deviation from a Gaussian PDF. In our study we have
quantified this and many other such deviations for all the field
increments $\delta \psi, \, \delta \phi, \, \delta u_{||}, \, \delta
b_{||}, \, \delta \omega$ , and $\delta j$.  We have quantified the
scale-dependent deviations of their PDFs from Gaussian ones by
computing hyperflatnesses. The most significant deviations occur
for the PDFs of $\delta \psi$ and $\delta \phi$; for these
increments we obtain their order-$p$ structure functions, in the
range $10 \leq l/l_{\rm inj} \leq 30$, and from these the multiscaling
exponent ratios $\zeta_p^\phi/\zeta_2^\phi$ and
$\zeta_p^\psi/\zeta_2^\psi$. Our exponent ratios suggest 
that, if there is any multiscaling at all, it is very mild;  
Ref.~\cite{celani} has 
suggested that $\zeta_2^\psi=1$. A decisive 
confirmation of multiscaling here requires very-high-resolution DNSs
that lie beyond our computational resources.

We have also obtained algebraically damped oscillations in
$S_2^{\omega}$ and similar but weaker ones in $S_2^{u_{||}}$. We
have shown how these damped oscillations can be understood
analytically. Once these oscillations are removed, or we move to length scales
$l$ where they have been damped, we find that that $S_p^{\omega}$
and $S_p^{u_{||}}$ approach constant values, i.e., they have 
a universal scaling exponent, which is $0$; such a universal exponent 
has been reported in some other turbulent systems with
inverse cascades~\cite{falkovich}. 

Our study has been restricted to a bare magnetic Prandtl number $Pr_M =1$
the effective Prandtl number $Pr_M=\nu_{\rm eff}/\eta_{\rm eff}$ is also
close to 1. Studies of our 2D MHD system at different 
$Pr_M$ will be presented elsewhere.

We hope our extensive study of the statistical properties of the
inverse-cascade regime in 2D MHD turbulence will stimulate
experimental studies of these properties.

\section{acknowledgements} 

We thank CSIR, UGC, and DST (India) for support, SERC (IISc) for
computational resources, and  A. Bhatnagar, A. Gupta, R. Majumder, and
S.S. Ray for discussions. 


\end{document}